%% file: main.tex
\documentclass[12pt]{iopart}

\usepackage{xurl}
\usepackage[usenames,dvipsnames,svgnames]{xcolor}
\usepackage{graphicx}

\expandafter\let\csname equation*\endcsname\relax
\expandafter\let\csname endequation*\endcsname\relax
\usepackage[cmex10]{amsmath} 
\usepackage{array}
\usepackage[pagewise]{lineno}
\usepackage{amssymb}
\usepackage[linesnumbered,ruled,vlined]{algorithm2e}
\usepackage{booktabs}
\usepackage[hidelinks]{hyperref}
\usepackage{caption}
\usepackage{placeins}
\usepackage{multirow}
\usepackage{arydshln} 
\usepackage{multicol}
\usepackage{changebar}
\newcommand{\R}{\mathbb{R}}
\DeclareMathOperator*{\argmin}{arg\,min}

\renewcommand{\S}{\mathbf{S}}
\newcommand{\seq}{\boldsymbol{\sigma}}
\renewcommand{\Im}{\text{Im}}

\newcommand{\CU}{\boldsymbol{\mathcal{U}}}
\newcommand{\CH}{\boldsymbol{\mathcal{H}}}
\newcommand{\CS}{\boldsymbol{\mathcal{S}}}
\newcommand{\wrapped}

\usepackage[sort]{natbib}
\setcitestyle{authoryear}

\begin{document}

\title{Optimization of array encoding for ultrasound imaging}

\author{Jacob Spainhour$^1$, Korben Smart$^2$, Stephen Becker$^1$ and Nick Bottenus$^3$}

\address{$^1$ Department of Applied Mathematics, University of Colorado Boulder, USA}
\address{$^2$ Department of Physics, University of Colorado Boulder, USA}
\address{$^3$ Paul M. Rady Department of Mechanical Engineering, University of Colorado Boulder, USA}
\eads{\mailto{Jacob.Spainhour@colorado.edu}, \mailto{Korben.Smart@colorado.edu}, \mailto{Stephen.Becker@colorado.edu}, \mailto{Nick.Bottenus@colorado.edu}}

\maketitle

\begin{abstract}
\input{abstract}  
\end{abstract}


\input{introduction}

\section{Methods}
\input{transmit_encoding_theory}
\input{beamforming_imaging_procedure}
\input{machine_learning_framework}

\input{training_data_acquisition}

\section{Results}
\input{basic_results}
\input{experimental_results}

\input{discussion}

\input{conclusions}

\newcommand{\newblock}{}
\bibliographystyle{dcu}

\bibliography{refs}

\end{document}

%% file: abstract.tex
\textit{Objective}: The transmit encoding model for synthetic aperture imaging is a robust and flexible framework for understanding the effects of acoustic transmission on ultrasound image reconstruction. 
Our objective is to use machine learning (ML) to construct scanning sequences, parameterized by time delays and apodization weights, that produce high-quality B-mode images.

\textit{Approach}: We use a custom ML model in PyTorch with simulated RF data from Field II to probe the space of possible encoding sequences for those that minimize a loss function that describes image quality.
This approach is made computationally feasible by a novel formulation of the derivative for delay-and-sum beamforming.

\textit{Main Results}: When trained for a specified experimental setting (imaging domain, hardware restrictions, etc.), our ML model produces optimized encoding sequences that, when deployed in the REFoCUS imaging framework, improve a number of standard quality metrics over conventional sequences including resolution, field of view, and contrast.
We demonstrate these results experimentally on both wire targets and a tissue-mimicking phantom.

\textit{Significance}: This work demonstrates that the set of commonly used encoding schemes represent only a narrow subset of those available.
Additionally, it demonstrates the value for ML tasks in synthetic transmit aperture imaging to consider the beamformer within the model, instead of purely as a post-processing step.

%% file: introduction.tex
\section{Introduction}
In classical ultrasound imaging, transmissions from a transducer array are focused towards specific locations in the target medium, and the backscattered echoes are collected and processed into an image that is clear at these locations~\citep{cobbold_foundations_2007}.
More modern systems use a synthetic transmit aperture system that combines the echoes from multiple transmissions to recreate this focus across multiple points in the region.
In the ideal case, data acquisition is performed by firing each array element individually in sequence and echoes are received by every element in parallel.
The full set of element-to-element signals is then combined into a B-mode image via post-processing that achieves focus even at depth~\citep{jensen_synthetic_2006}. 
However, the large number of transmissions needed paired with the relatively low power of each means that, when performed directly, this procedure creates images with poor frame rate and low signal-to-noise ratio (SNR), making it impractical for clinical application~\citep{chiao_sparse_1997}.

Retrospective Encoding For Conventional Ultrasound Sequences, or REFoCUS, is an alternative imaging framework that considers the responses from an \textit{arbitrary} transmission as a linear combination of these pairwise element responses~\citep{bottenus-recovery-2018, ali-extending-2020}.
In this way, the transmission sequence described by time delays and apodization weights parameterizes an \textit{encoding} of the ground-truth basis of element-to-element signals, the multistatic (STA) data set.
By using the specific sequence used to acquire data, the echoed response signals can be \textit{decoded} to produce a robust approximation of this basis.
This approximation can then be focused throughout the imaging domain as post-processing, making a B-mode image that achieves the desired SNR and focal depth.
Conventional transmission sequences are often selected according to geometric principles (e.g., focused, planewave, or diverging beams) or according to spatial codes (e.g., Hadamard~\citep{chiao_sparse_1997} or S-sequence~\citep{harrison_s-sequence_2014}).
However, the REFoCUS framework generalizes the encoding framweork to allow for a uniform treatment of other categories of transmissions~\citep{bottenus-encoding-comparison-2023}.

The quality of the resultant B-mode image---measured in terms of resolution, field of view (FOV), and artifacts---is highly dependent on properties of the transmit sequence used for data acquisition.
As an example of the importance of beam geometry, focused beams provide very high resolution, but only around a particular focal point.
Another influential property is simply the number of transmissions.
While the number of transmissions ideally matches or exceeds the number of array elements, practical frame rate considerations often restrict this.
In the resulting underdetermined case, there is active research that seeks to minimize the loss of information during encoding and decoding.
Some groups have shown success using randomly assigned delays, as such schemes lead to a very well conditioned encoding, in the sense that the relevant encoding matrices are full rank and therefore stably invertible~\citep{you_pixel-oriented_2021}.
Similarly, a spatial code for element apodizations can be applied to produce a lossless encoding (or near-lossless when the matrix must be truncated~\citep{zhang_partial_2021}), which in turn increases SNR during imaging.

However, the space of all possible encoding sequences dwarfs those with these sorts of immediately clear and desirable properties~\citep{spainhour-proceedings-2022}.
In this paper, we present a novel machine learning (ML) model for data acquisition and imaging which we train to efficiently search the set of all possible transmission sequences, identifying those that lead to high-quality images using the REFoCUS encoding framework.
Notably, our training workflow is designed to measure the quality of a sequence after image formation, better reflecting how performance is judged in a clinical setting.
Furthermore, the only trainable parameters of this machine learning model are a physical description of the transmission sequence, such that a trained ML model is parameterized by an optimized encoding sequence.
This optimized encoding sequence can then be considered independently of the ML model that generated it, allowing it to be deployed classically within the REFoCUS framework to achieve image quality beyond what is attainable with conventional sequences.

\subsection{Related Work}

Techniques in machine learning have gained a significant amount of research attention in the field of ultrasound imaging, particularly in the application of deep neural networks (DNNs)~\citep{liu-deep-learning-2019, hyun-cubdl-2021}.
Following the unprecedented successes of the deep convolutional neural network (CNN) AlexNet in non-medical imaging tasks~\citep{krizhevsky-alexnet-2017}, similar networks are frequently trained and utilized for ultrasound-adjacent tasks as an additional post-processing step to a conventional imaging pipeline by performing classification, segmentation, etc.\ on image data~\citep{mischi-deep-learning-2020}.

Other deep learning approaches perform image formation directly.
In a prototypical setup, a DNN is fed the received echoes of a transducer array to produce a B-mode image, aiding or even supplanting the use of a conventional beamformer~\citep{sloun-deep-learning-2020, goudarzi-deep-reconstruction-2022}.
Often the goal is to create a network that can be evaluated faster than classical imaging techniques without degrading image quality~\citep{chen_apodnet_2021}.
Other times, the DNN is designed to directly \textit{improve} image quality.
For example,~\cite{hyun-beamforming-NN-2019} utilize a CNN as a beamformer that reduces speckle, citing considerable advantages over conventional delay-and-sum techniques.

While a well-trained DNN can offer these and other benefits, their use necessarily introduces complications.
Most notably, neural networks are susceptible to hallucinations, in which it unexpectedly generates features that are not present in the underlying data, an issue that is uniquely consequential in medical imaging~\citep{bhadra-hallucinations-2021}.
Our approach to machine learning entirely circumvents this and other issues by using a custom architecture whose forward operation is entirely acoustically motivated. 
Specifically, the proposed ML method improves imaging quality \textit{exclusively} through an optimized parameterization of experimental hardware via the transmit sequence. 
After data is acquired with an optimized sequence, the B-mode image to be evaluated is generated through the REFoCUS imaging framework, which itself is based solely on the linear nature pulse-echo responses (We describe the decoding and imaging processes in detail in Section~\ref{sec:encoding_theory} and Section~\ref{sec:beamforming_and_imaging} respectively).

A similar strategy to our own is employed by~\cite{chen_apodnet_2021}, in which a machine learning model is trained to optimize the apodization weights of planewave transmissions, although a DNN is used to decode the received echoes.
Specifically, a stacked denoising autoencoder architecture (a type of DNN whose first trainable layer is taken to be the weights themselves) is trained to optimize the reconstruction of the multistatic data set.
In contrast, our ML model allows full control of the encoding sequence through a continuous treatment of apodization weights \textit{and} time delays, with no other trainable parameters.
At the same time, our training workflow optimizes the encoding sequence according to direct measures of image quality.
In doing so, we provide notable counterexamples to the principle that a better reconstruction of the multistatic data set leads to higher quality images. 

By detaching our machine learning model from conventional neural network architectures, our approach obtains theoretical and practical advantages beyond avoiding hallucinations.
First, an optimized encoding sequence generated by our ML model can be analyzed separately from the model itself, such that its beneficial features can be identified and further studied.
This is in contrast to a trained DNN, whose per-layer weights and biases have no meaning outside the context of the DNN.
Second, the smaller parameter space of our ML model dramatically reduces the training time needed to generate an optimized encoding sequence.
This allows for greater flexibility in the overall ML workflow, as a different configuration of the ML model can be retrained at a minimal cost when the experimental setting changes (e.g.,\ targeting a different viewing depth or prioritizing different image features via an alternate loss function).
Finally, optimization of data acquisition through the transmit sequence makes our imaging process very robust to different targets.
While conventional DNNs are typically trained for very specific imaging tasks on a constrained set of training data, we will show that the proposed approach not only generalizes well to out-of-sample training data, but also to data with dramatically different features.
This is because our ML model is fundamentally optimized based on the underlying acoustic principles of ultrasound imaging, where the processing of backscattered echoes is influenced primarily through manipulation of the transmission sequence.

%% file: transmit_encoding_theory.tex
\subsection{Transmit Encoding Theory}\label{sec:encoding_theory}
We consider a linear transducer array of $N_E$ physical elements, of which $N_T$ are capable of transmitting a diverging wave and $N_R$ can receive backscattered echoes.
Moreover, it is typical to have elements that perform both functions, such that $N_T = N_R = N_E$. 
Taken across all pairs of transmit and receive elements, collected RF (radio frequency) signals form a time dependent $N_{T} \times N_{R}$ matrix $\mathbf{U}(t)$, within which the matrix element~$u_{TR}(t)$ is the signal observed by transmitting on element $T$ and receiving on element $R$.
In the synthetic transmit aperture model, the complete collection of pulse-echo response signals between pairs of transmit and receive elements make up the multistatic data set.
Because these signals can be delayed and summed across the receive channel to produce transmit focus throughout the domain, as in delay-and-sum (DAS) beamforming, these data can be considered the mathematical basis of our imaging~\citep{jensen_synthetic_2006}.

As previously discussed, it is inadvisable to construct $\mathbf{U}(t)$ directly by firing each transmit element in sequence across a total of $N_T$ transmissions~\citep{chiao_sparse_1997}.
Instead, the REFoCUS technique seeks to construct an approximation of $\mathbf{U}(t)$ using the responses from an arbitrary scanning sequence.
We consider a scanning sequence of $N_M$ separate transmissions, each of which is parameterized by $N_T$ time delays $t_{MT}$ and apodization weights $w_{MT}$.
The response observed by receive element $R$ from transmit $M$ can be described as a weighted sum of time delayed matrix elements from the multistatic data set,
\begin{align}
    s_{MR}(t) = \sum_{T=1}^{N_T} w_{MT}u_{TR}(t - t_{MT})\,,
\end{align}
which are similarly collected into a time dependent $N_M \times N_R$ matrix $\mathbf{S}(t)$ of focused responses.
Collected across each transmit, these time delays and apodization weights make up the \textit{encoding sequence}, denoted $\seq = (t, w)_{MT}$.

It is now convenient to work in the frequency domain, where each time delay of~$u_{TR}(t)$ becomes a complex phase shift of $u_{TR}(\omega)$~\citep{ali-extending-2020}.
For simplicity, we represent the frequency dependent Fourier transform of $\mathbf{U}(t)$ and $\mathbf{S}(t)$ as $\mathbf{U}(\omega)$ and $\mathbf{S}(\omega)$ respectively.
In doing so, we can express the transmission response of receive element~$R$ to transmission $M$ as
\begin{align}
    s_{MR}(\omega) = \sum_{T=1}^{N_T} w_{MT}\exp(- j \omega t_{MT})u_{TR}(\omega)\,.
\end{align}
This describes a linear relationship between the multistatic data set $\mathbf{U}(\omega)$ and the transmission responses $\mathbf{S}(\omega)$ that is fully parameterized by $\seq$, given by
\begin{align}
    \mathbf{S}(\omega) = \mathbf{H}(\omega)\mathbf{U}(\omega)\,,
\end{align}
where $\mathbf{H}(\omega)$ is a frequency dependent $N_M \times N_T$ \textit{encoding matrix} given by

\begin{equation}
	\mathbf{H}(\omega)=\begin{bmatrix}
		w_{1,1}\exp(-j\omega t_{1,1}) & \dots & w_{1,T}\exp( j\omega t_{1,T})\\
		w_{2,1}\exp(-j\omega t_{2,1}) & \dots & w_{2,T}\exp( j\omega t_{2,T}) \\
		\vdots & \ddots & \vdots\\
		w_{M,1}\exp(-j\omega t_{M,1}) & \dots & w_{M,T}\exp( j\omega t_{M,T})\\
	\end{bmatrix}\,.
	\label{eqn:H}
\end{equation}
In practice, we collect only discrete, equispaced samples from $\mathbf{S}(t)$, from which we have a total of $N_\omega$ angular frequencies $\omega$. 
Because calculations involving the collection of matrices $\mathbf{S}(\omega)$ are typically performed in parallel across frequencies, we use calligraphic letters to denote the collection of all frequencies.
As an example, we denote $\CS_\omega = \mathbf{S}(\omega)$, and represent the above equation more compactly as $\CS = \CH \CU$. 
When we wish to emphasize the dependence of our encoding matrices $\CH$ on the sequence $\seq$, we add $\seq$ as a subscript.

The objective of the REFoCUS framework is to create an approximation of the multistatic data set $\widehat{\CU}$ from the recorded echoes in $\CS$.
This is done by applying a frequency dependent \textit{decoder} $\mathbf{H}^\dagger(\omega)$ to $\mathbf{S}(\omega)$, resulting in 

\begin{equation}
	\widehat{\CU}=\CH^\dagger\CS=(\CH^\dagger\CH)\CU\,.
	\label{eqn:uhat}
\end{equation}

This decoder is theoretically arbitrary, with options explored in the literature ranging from a per-frequency conjugate transpose~\citep{bottenus-recovery-2018} to a fully trained neural network~\citep{chen_apodnet_2021}. 
In this work, we follow~\cite{ali-extending-2020, bottenus-encoding-comparison-2023} and apply a Tikhonov regularized pseudoinverse that has been shown to work well experimentally.
This choice of decoder depends only on the original encoding matrix, which is itself dependent only on our sequence $\seq$.
The Tikhonov decoder is given explicitly by
\begin{equation}
    \CH^\dagger=(\CH^*\CH+\gamma^2 \boldsymbol{\mathcal{I}}_{N_E})^{-1}\CH^*\,,
\end{equation}
where $\CH^*$ is the conjugate transpose of $\CH$ and $\gamma$ is a regularization constant.
This regularization both suppresses the contribution of noise to the recovered multistatic channel data and produces a more stable inversion of $\CH$, which together results in a more accurate and robust recovery of $\widehat{\CU}$.
This is particularly important in the underdetermined system of interest, in which the number of transmits $N_M$ is significantly smaller than the number of array elements $N_E$.
In all usages of regularization herein we take a value of $\gamma = 0.1\sigma_{\text{max}}(\omega)$, where $\sigma_{\text{max}}(\omega)$ is the spectral norm of each per-frequency encoding matrix $\mathbf{H}(\omega)$.
This particular value has been observed empirically to avoid issues of over- or under-regularization in the presence of noise~\citep{bottenus-encoding-comparison-2023}. 

%% file: beamforming_imaging_procedure.tex
\subsection{Beamforming and Imaging Procedure}\label{sec:beamforming_and_imaging}
Given a multistatic data set $\CU$, we produce the 2D B-mode image of interest with conventional DAS beamforming.
This technique maps ground-truth or approximated multistatic IQ data (in-phase and quadrature data derived from the experimentally acquired RF data) at each frequency to a single IQ data matrix, taking a sum of the time delayed data in $\widehat{\CU}$. 
By manipulating these post-acquisition time delays, an image is created that is focused at individual pixels in the imaging domain.
Mathematically, we denote this beamforming operation as $\mathcal{B} : \mathbb{C}^{N_M \times N_R \times N_\omega} \to \mathbb{C}^{N_x \times N_z}$, where $(N_x, N_z)$ are the number of pixels in the lateral and axial directions of our image. 

To make the resulting image suitable for interpretation by a human observer, we apply a number of non-linear post-processing steps to this re-focused IQ data.
These include the computation of the signal envelope, log-scaling the dynamic range, and clipping the image to a minimum decibel value.
In our notation, we represent these operations collectively as~$|\cdot|_{\Im} :~\mathbb{C}^{N_x \times N_z} \to \mathbb{R}^{N_x \times N_z}$. 
When applied in the experimental context, the total forward operation of our ML model takes in transmission responses collected in $\CS$ and an encoding sequence $\seq$, and produces an image represented by the real matrix $|\mathcal{B}( \CH_{\seq}^\dagger \CS)|_{\Im}$.

To ensure that the generated encoding sequences is useful in a clinical setting, we train our ML model with a loss function $\mathcal{L}:~\R^{N_x \times N_z}\to~\R^+$ that directly measures image quality.
While there is flexibility in the specific choice of loss function~\citep{hyun_optimization_2022}, achieving a lower value for the loss should correspond to higher resolution, higher SNR, wider FOV, fewer artifacts, etc.

A clear choice is to use a loss function that quantifies these qualities directly.
For example, the generalized contrast to noise ratio (gCNR) is a measure of histogram overlap between the brightness of predefined speckle and anechoic regions, such that a higher value indicates greater contrast in the image~\citep{rodriguez-molares_generalized_2019}. 
We can in principle improve the gCNR by minimizing the loss function 
\begin{align}
    \mathcal{L}_{\text{gCNR}}(\widehat{\mathbf{X}}) := 1 - \text{gCNR}(\widehat{\mathbf{X}})\,,
\end{align}
However, complications with implementation make this loss function impractical.
In particular, the histogram operator is not differentiable and standard continuous approximations are unstable, which makes the metric incompatible with backpropagation during training.

On the other hand, a natural and easily implementable choice is a normalized $\ell^2$ comparison to a reference image that has desirable qualities.
This introduces further flexibility in the choice of imaging target, which we discuss in Section~\ref{sec:acquisition_training_data} and Section~\ref{sec:imaging_target}.
With a given image~$\widehat{\mathbf{X}}$ and reference image~$\mathbf{X}$, this loss is then defined by
\begin{align}\label{eqn:l2loss}
    \mathcal{L}_{\ell^2}(\widehat{\mathbf{X}}; \mathbf{X}) := \frac{1}{N_xN_z}
    \big\|
    \mathbf{X} - \widehat{\mathbf{X}}
    \big\|^2_2\,.
\end{align}
Importantly, we will show that training the ML model to minimize $\mathcal{L}_{\ell^2}$ still makes a meaningful improvement to the gCNR of produced images.

%% file: machine_learning_framework.tex
\subsection{Implementation of the Proposed Machine Learning Model}\label{sec:implementation}
We define an optimized encoding sequence $\seq^*$ as one that, for a given loss function~$\mathcal{L}$, beamformer $\mathcal{B}$, and decoder $\CH^\dagger_{\seq}$, minimizes the non-convex optimization problem
\begin{align}
    \min_{\seq \in \Sigma} \mathbb{E}\;\mathcal{L}\left(|\mathcal{B}( \CH_{\seq}^\dagger \CS)|_{\Im}\right)\,,
\end{align}
where this expectation is taken over all possible transmission responses $\CS$.
As it is impossible to evaluate this expectation directly, we instead approach the minimization problem from a machine learning perspective, where we instead minimize over a representative sample of training data $\{\CU_i\}$.
Each piece of ground-truth multistatic data is related to a transmission response by our encoding matrix~$\CH_{\seq}$ by $\CS_i = \CH_{\seq} \CU_i$.

Similarly, we must restrict the space of encoding sequences to those that are practically realizable on a physical system, which we denote as $\Sigma$.
For example, we limit the total transmission power by restricting the set of apodization weights $\{w_{MT}\}$ to an $\ell^\infty$ ball, effectively clipping each value to the range $[-1, 1]$.
Other restrictions stem from quantization of numerical values to match machine clock cycles.
These small discretizations are much more dependent on the specific acquisition system, but also less impactful to the optimized result. 

Altogether, we are left to solve

\begin{align}
    \label{eqn:optimized_sequence}
    \seq^* = \argmin_{\seq \in \Sigma} \frac{1}{N_U} \sum_{i=1}^{N_U} \mathcal{L}\left(|\mathcal{B}( \CH_{\seq}^\dagger \CH_{\seq} \CU_i)|_{\Im}\right)\,. 
\end{align}

\begin{figure}[tbp]
  \centering
  \includegraphics[width=\textwidth]{ 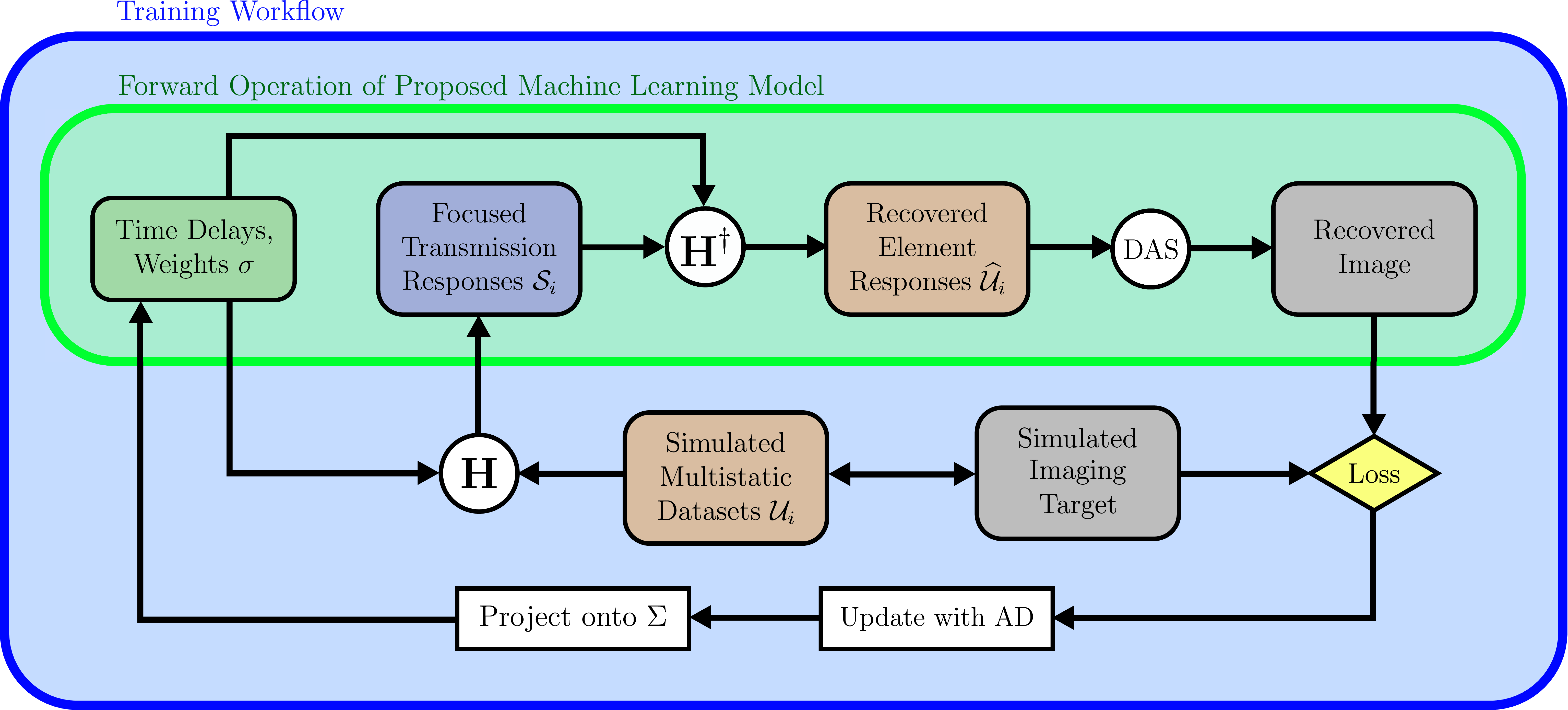}
  \captionsetup{font=small}
  \caption{The training procedure for the proposed ML model for data acquisition and imaging, which is fully parameterized by an encoding sequence. Simulated multistatic training data is encoded and decoded according to our model parameters, and an image is formed using delay-and-sum beamforming. The resulting B-mode image is evaluated by comparison to a target image, or with some other data obtained during the simulation, i.e.,\ target position.}
  \label{fig:flowchart}
\end{figure}

Numerically, we implement and train the ML model that minimizes Equation~\ref{eqn:optimized_sequence} in PyTorch, a Python package for building deep learning and other machine learning models~\citep{paszke-pytorch-2019}.
This platform is flexible to our unique architecture, which lacks the abstract layers of a neural network and instead propagates information only through acoustically motivated operations, namely the encoding/decoding of the multistatic data and the generation of a B-mode image with DAS beamforming.
In all other respects, we mirror the conventional ML pipeline by minimizing our loss function over batches of training data.
We perform the numerical optimization using the Adam algorithm, an adaptive version of standard stochastic gradient descent~\citep{kingma-adam-2017}, and accommodate the constraint set~$\Sigma$ for physically realizable encoding sequences with a simple projection of our sequence during each update.
The update loop of our ML model is depicted in Figure~\ref{fig:flowchart}, for which we emphasize the portion that is deployed in an experimental context.

To effectively update the parameters of any ML model requires knowledge of first order derivatives, which we obtain using the reverse-mode automatic differentiation, or \textit{backpropagation}, capabilities of PyTorch~\citep{griewank-autodiff-2008}.
In the context of ultrasound imaging, standard techniques are often sufficient when the loss function depends primarily on multistatic data~\citep{chen_apodnet_2021} or on individual pieces of RF data~\citep{noda-shapeestimation-2020}.
This is because platforms like PyTorch have built-in analytic derivatives for operations that are common in deep learning applications, such as matrix multiplication and Fourier transforms.
Gradients for all other operations can be constructed at runtime at the cost of additional computation time and memory overhead.
However, this process is made difficult when an ML model directly incorporates beamforming, which we consider a critical component to produce high-quality encoding sequences (See Section~\ref{sec:loss_function}).
Backpropagation through the DAS beamformer rapidly becomes a computational bottleneck when it is implemented directly as a sum of linear interpolations of IQ data into per-pixel focused channel data, as the number of intermediate values that must be stored in memory grows rapidly.
This is especially relevant while training the proposed ML model, as we must simultaneously perform this expensive calculation over entire batches of training data.

In our application, we circumvent this issue through a novel PyTorch implementation of the derivative of the DAS beamforming operator $\mathcal{B}$.
Because the beamformed image $\mathcal{I}$ is linear as a function of the multistatic data $\mathcal{U}$~\citep{perrot_das_2021}, the analytic derivative needed for backpropagation is simply its adjoint operator.
We provide PyTorch the DAS adjoint for arbitrary data, which eliminates the additional cost and memory overhead.

We do not directly construct and transpose the DAS beamforming matrix, as it is impractically large for reasonably sized multistatic data, even under a sparse representation~\citep{perrot_das_2021}.
Instead, we apply this adjoint matrix-free, recognizing that $\mathcal{B}$ can be described as a sum of linear interpolations, or
\begin{align*}
    \mathcal{B}(\CU) = \text{Sum}(\text{Interpolate}(\CU))\,.
\end{align*}
From this, we derive from simple analytic formulas that
\begin{align*}
    \mathcal{B}^*(\mathcal{I}) = \text{Interpolate}^*(\text{Sum}^*(\mathcal{I}))\,.
\end{align*}
These two component adjoints have known, albeit obscure formulas~\citep{claerbout-adjoints-2014}, and so we provide pseudocode for both operations for clarity.
The code used to perform the optimization, along with the RF data that support the findings of this study, are available at \href{http://github.com/jcs15c/optimal_ultrasound_encoding}{\url{github.com/jcs15c/optimal\_ultrasound\_encoding}}~\citep{spainhour-code-2024}.

\begin{algorithm}[htbp]
    \caption{DAS Beamformer $\mathcal{B}$}\label{alg:DAS}
    \DontPrintSemicolon
    \KwIn{$\CU$: Multistatic Data}
    \KwOut{$\mathcal{I}$: Beamformed Image}
    Initialize focused IQ data at each pixel to zero\;
    \For{each transmit-receive element pair}{
        \tcc{Linearly interpolate per-pixel time delays onto the channel data for the transmit-receive element pair}\;
        \For{each pixel in the image}{
            Identify the pair of IQ data the pixel focused time delay is between.\;
            Add a linear combination of these two IQ data to the focused IQ data \;
        }
    }
\end{algorithm}

\begin{algorithm}[htbp]
    \caption{DAS Beamformer Adjoint $\mathcal{B}^*$}\label{alg:DAS-adjoint}
    \DontPrintSemicolon
    \KwIn{$\widetilde{\mathcal{I}}$: Data of the same shape as $\mathcal{I}$}
    \KwOut{$\widetilde{\mathcal{\CU}}$: Data of the same shape as $\CU$}

    Initialize output to zeros\;
    \For{each transmit-receive element pair}{
        \tcc{Perform adjoint-interpolation of $\widetilde{\mathcal{I}}$ based on per-pixel time delays}\;
        \For{each pixel in the image}{
            Identify pair of IQ data the pixel focused time delay is between\;
            To each index of the output where the closest pair is located, add the corresponding value of the linear combination in Algorithm~\ref{alg:DAS} 
        }
    }
\end{algorithm}

%% file: training_data_acquisition.tex
\subsection{Acquisition of Training and Testing Data}\label{sec:acquisition_training_data}

\begin{table}[b]
	\centering
    \begin{tabular}{l r}\toprule
		Parameters & Simulated array\\\midrule
		Element count & 64\\
		Element pitch & 0.3 mm\\
		Element kerf & 0.01 mm\\
		Center frequency & 3 MHz\\
		Bandwidth & 70\%\\
		\bottomrule
	\end{tabular}
 \captionsetup{font=small}
	\caption{Simulated Field II Transducer Parameters}
	\label{tab:simtransducers}
\end{table}

While training on true \textit{in vivo} data would best reflect the experimental setting, it is understood that large collections of general purpose, labeled clinical data are rare~\citep{liu-deep-learning-2019}.
Instead, we utilize simulated multistatic data created with Field~II~\citep{jensen_field_2004} to train and evaluate our proposed ML model.
The simulated transducer configuration is described in Table~\ref{tab:simtransducers}. 
We have found that multistatic data depicting randomly placed anechoic lesions in an underdeveloped speckle pattern is a particularly effective class of training data, as encoding sequences trained on it generalize well to other classes of data (See Section~\ref{sec:significance_training_data}).

For training, we generate with Field~II a collection of 500 instances of underdeveloped speckle data, of which 20\% form a validation set for by-hand tuning of our (reasonably few) hyperparameters.
Notably, the proposed model is exposed \textit{only} to this type of data during training.
For thorough testing of the optimized encoding sequence, we use additional samples of underdeveloped speckle data, as well as collections of other types of simulated data.
These include conventional imaging targets, namely isolated point scatterers and anechoic lesions in fully developed background speckle.
These data are also used to evaluate the optimized encoding sequence according to standard ultrasound imaging quality metrics, such as the cystic resolution and gCNR respectively.

To explore performance on more complex RF data, we follow a procedure similar to~\cite{hyun-beamforming-NN-2019} and generate arrangements of scatterers with amplitudes weighted according a grayscale image.
These images are derived from a set of 160 samples from the publicly available validation set of an ImageNet competition~\citep{howard-imagenet-2018}, where we have extracted and smoothed the middle 256$\times$256 pixels from each and converted them to grayscale.
The first type of image-derived data interpolates the pixel values of each grayscale image to nearby scatterers that are distributed throughout the spatial domain, allowing us to consider data with scatterers of arbitrary amplitude.
The second type of image-derived completely removes scatterers near pixels of sufficiently low brightness, resulting in anechoic regions in background speckle that are more arbitrary than standard circular cysts.

We also test our optimized encoding sequences on experimental hardware with both a tissue-mimicking phantom and a wire target, as described in Section~\ref{sec:experimental}.

We must similarly consider the imaging target used by the loss function in Equation~\ref{eqn:l2loss} during the supervised learning.
A natural choice is an ``unencoded'' image, where each simulated ground-truth multistatic data set is directly focused with DAS beamforming, thereby removing artifacts induced by encoding.
Alternatively, one can compare to a more ``idealized'' target that represents the ground-truth \textit{contrast}, as this also reduces the influence of any particular speckle pattern during training.
Each image of ground-truth contrast is generated alongside the associated simulated multistatic data, where spatial masks are used to create homogeneous anechoic and scattering regions.
An alternative approach to describing ground-truth echogenicity would be to directly convolve a high-quality point spread function (PSF) with each scatterer~\citep{goudarzi-deep-reconstruction-2022}, but we find our approach using a spatial mask to be simpler computationally.
For image-derived data, the ground-truth contrast is simply the original grayscale image.
During training, we use ground-truth contrast targets, and discuss the consequences of this decision in Section~\ref{sec:imaging_target}.

\begin{figure}[tbp]
  \centering
  \includegraphics[width=\textwidth]{ 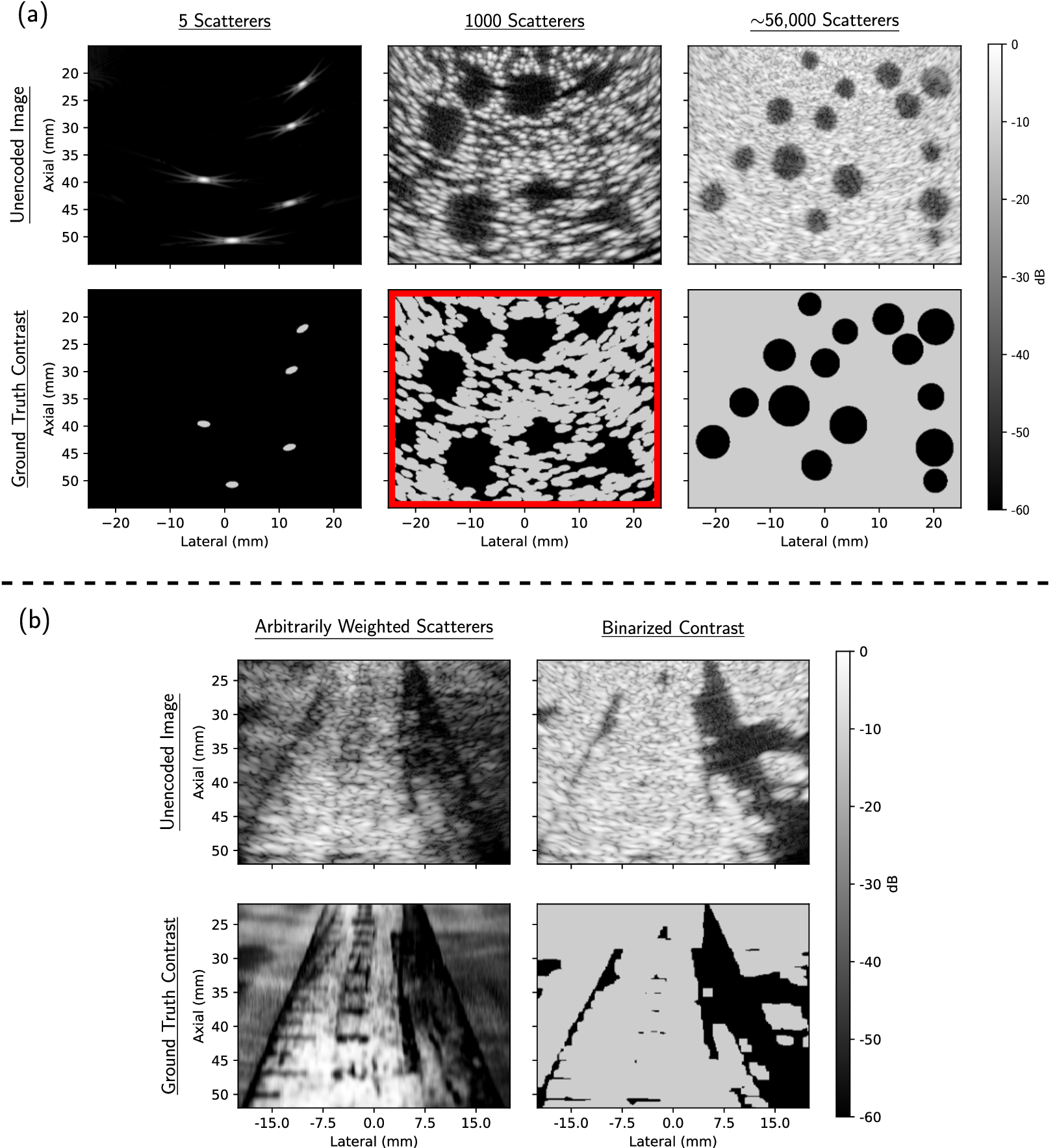}
  \captionsetup{font=small}
  \caption{The types of simulated data and imaging target type on which the ML model is trained and/or evaluated. (a) Conventional imaging targets generated from the uniform responses of individual scatterers in an anechoic field. (b) Imaging targets for image-derived data, where the amplitude of each scatterer is weighted according to a grayscale image. While all categories are used for validation of the ML model, numerical experiments suggest that training on the class of data emphasized in red (ground-truth contrast, underdeveloped speckle) produces the best optimized encoding sequences.}
  \label{fig:training_data}
\end{figure}

In Figure~\ref{fig:training_data} we show an example of each target type for each category of simulated data, highlighting the particular configuration of data that is exclusively used during training.

%% file: basic_results.tex
\subsection{Optimization of Both Time Delays and Apodization Weights}\label{sec:both}

We now demonstrate our capability to effectively train the proposed ML model, thereby producing encoding sequences that meaningfully improve image quality.
In the most general case, the available hardware allows us to manipulate both time delays and apodization weights for each array element. 
This allows the greatest freedom when designing encoding sequences, and predictably the greatest improvement over conventional sequences.
As a prototypical configuration of the training workflow, we train the ML model using 400 simulated instances of the previously described underdeveloped speckle background with anechoic lesions, processed in batches of size 8.
We evaluate the model with the $\ell^2$ loss function in Equation~\ref{eqn:l2loss} using ground-truth contrast as the imaging target.
The training is performed using the Adam optimizer using a learning rate of 0.1 over 25 epochs.
The available Verasonics hardware is limited to a minimum duty cycle of two clock cycles, which effectively nullifies low values for our weights.
As a result, we appropriately restrict the parameter space of encoding sequences $\Sigma$ during training (See Section~\ref{sec:implementation}).

While this is one of many possible configurations of the proposed training workflow (e.g., the choice loss function can vary according to the specific imaging task) we will see that the optimized encoding sequence created in the above configuration improves image quality fairly broadly across a number of different metrics.

\begin{figure}[htbp]
  \centering
  \includegraphics[width=\textwidth]{ 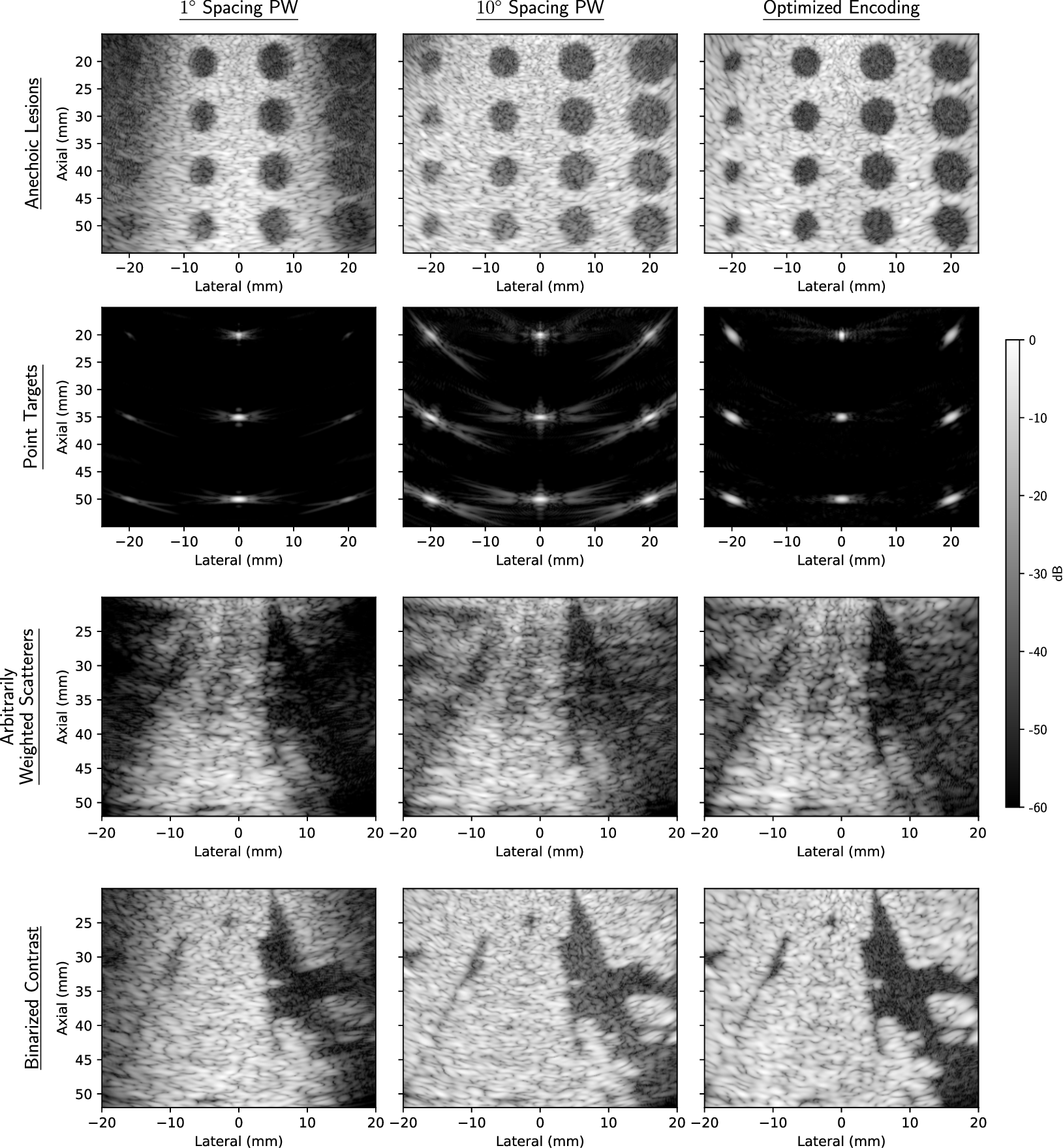}
  \captionsetup{font=small}
  \caption{Three choices for 15-transmit encodings applied to different imaging targets. We compare to imaging using planewaves with 1 degree of separation (left), planewaves with 10 degrees of separation (middle), and our novel optimized sequence (right). In all types of simulated data, we see considerable improvements in contrast and resolution over the conventional transmit encodings.}
  \label{fig:total_improvements}
\end{figure}

We first demonstrate these improvements by evaluating the optimized encoding sequence on several types of simulated ultrasound data in Figure~\ref{fig:total_improvements}, for which we uniformly observe significantly improved contrast and reduction of scattering artifacts.
For comparison, we use two conventional planewave encodings of varying maximum angle extent, as such sequences characterize a common tradeoff between FOV and resolution at depth~\citep{bottenus-encoding-comparison-2023}.
For example, the first generates 15 beams with 1 degree of separation, which results in a higher resolution along a narrower FOV.
This is particularly notable when imaging point targets, where some scatterers are essentially undetectable.
Steering these 15 planewaves sufficiently far apart to cover the entire imaging domain results in shallower contrast for anechoic regions, as there is more significant scattering throughout the domain.

By contrast, the proposed ML model has generated a sequence that recovers this wider FOV simultaneously with improved contrast.
Importantly, we observe through the improvements on the image-derived data that the optimized encoding sequence is performant regardless of the specific arrangement of scatterers in the field.
Instead, the point spread function for each scatterer is improved more generally and consistently throughout the FOV.

We show these improvements quantitatively in Table~\ref{tab:improvement_table}, where this prototypical optimized encoding sequence leads to improved $\ell^2$ error against the ground-truth contrast.
This is the same metric that is minimized over the training set of anechoic lesions in underdeveloped speckle (See Equation~\ref{eqn:l2loss}), and it is therefore natural that our ML model improves this metric for this type of data.
However, we see that these improvements in the $\ell^2$ loss are consistent across several \textit{other} types of imaging targets that are visually quite distinct.

\begin{table}[htbp]
    \centering
    \begin{tabular}{r c c c c}
        \toprule
        \multicolumn{5}{c}{Average $\ell^2$ Loss against Ground-Truth Contrast Map}\\
        \toprule 
                                                  & $1^\circ$ Spacing      & Full FOV Span          & Truncated              & Optimized                       \\
         \multicolumn{1}{c}{\textbf{Target Type}} & Planewaves             & Planewaves             & Hadamard               & Encoding                        \\\cmidrule(r){1-1} \cmidrule(l){2-5}
         Isolated                                 & \multirow{2}{*}{12.55} & \multirow{2}{*}{20.03} & \multirow{2}{*}{33.87} & \multirow{2}{*}{\textbf{8.12}} \\
         Point Targets                            &                        &                        &                        &                                 \\\cmidrule(r){1-1}
         Anechoic Lesions in                      & \multirow{2}{*}{583.9} & \multirow{2}{*}{331.6} & \multirow{2}{*}{366.3} & \multirow{2}{*}{\textbf{252.9}} \\
         Underdeveloped Speckle                   &                        &                        &                        &                                 \\\cmidrule(r){1-1}
         \multirow{2}{*}{Anechoic Lesions}        & \multirow{2}{*}{530.8} & \multirow{2}{*}{250.5} & \multirow{2}{*}{257.3} & \multirow{2}{*}{\textbf{205.2}} \\
                                                  &                        &                        &                        &                                 \\\cmidrule(r){1-1}
         Image-Derived                            & \multirow{2}{*}{256.6} & \multirow{2}{*}{166.8} & \multirow{2}{*}{174.4} & \multirow{2}{*}{\textbf{160.8}} \\
         Contrast                                 &                        &                        &                        &                                 \\\cmidrule(r){1-1}
         Binarized Image-                         & \multirow{2}{*}{369.9} & \multirow{2}{*}{258.8} & \multirow{2}{*}{274.7} & \multirow{2}{*}{\textbf{230.1}} \\
         Derived Contrast                         &                        &                        &                        &                                 \\
         \bottomrule
    \end{tabular}
    \captionsetup{font=small}
    \caption{We compute the average $\ell^2$ loss against the ground-truth contrast across 50 pieces of simulated data outside the training set. Although the encoding sequence is trained using only the data set of anechoic lesions in underdeveloped speckle, this metric is improved for each other type of data, emphasized in bold.}
    \label{tab:improvement_table}
\end{table}

We can also demonstrate that improvements in this $\ell^2$ loss are strongly associated with improvements in more conventional image quality metrics, namely the gCNR and cystic resolution, thereby justifying the decision to select an $\ell^2$ loss for training of the ML model.

\begin{figure}[htbp] 
  \centering
  \includegraphics[width=\textwidth]{ 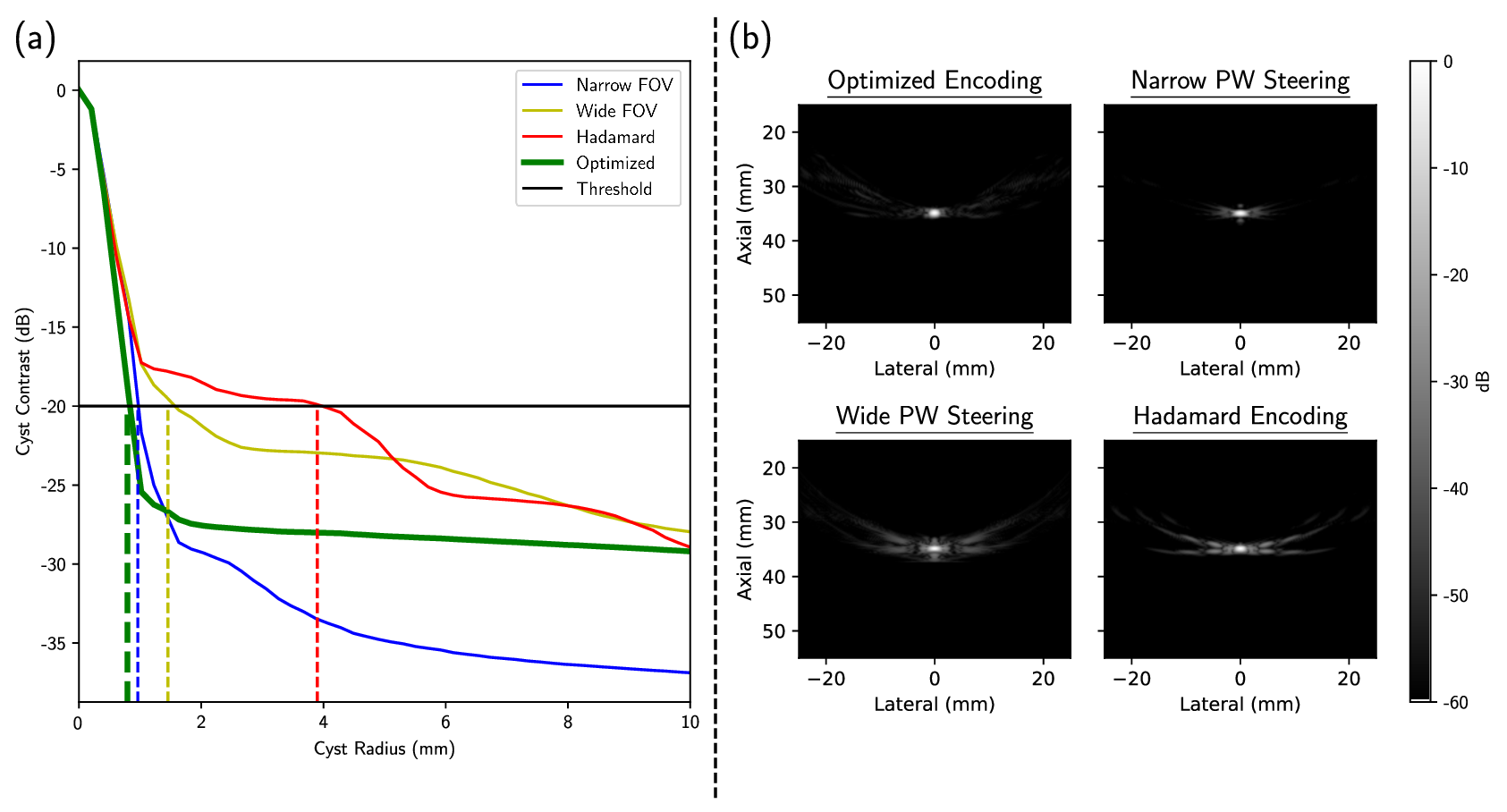}
  \captionsetup{font=small}
  \caption{(a) The contrast of an anechoic cyst as a function of cyst size, with a vertical line indicating the value of the cystic resolution. (b) The centered point targets for which the cystic resolution is measured. From this, we can see that the optimized encoding results in cystic resolution  comparable to that of narrow span planewaves.}
  \label{fig:cystic_resolution_explanation}
\end{figure}

The cystic resolution quantifies detectability of an anechoic cyst in background speckle with a given level of contrast~\citep{ranganathan-cystic-resolution-2007}.
We compute this metric through cystic \textit{contrast}, which is the concentration of energy within a certain distance from a point target in an anechoic field.
When cystic contrast is considered as a function of target radius, the cystic resolution is the minimum radius that achieves a desired level of contrast.
For our examples we use a contrast of -20 dB, a common threshold for lesion detectability that represents an order of magnitude difference from the surrounding speckle.
By explicitly plotting the cystic contrast for each encoding sequence in Figure~\ref{fig:cystic_resolution_explanation}(a), we can see that the optimized encoding sequence produces lesion detectability on par with narrowly concentrated planewaves, and that this conclusion is not sensitive to the exact choice of threshold.

\begin{figure}[htbp] 
  \centering
  \includegraphics[width=\textwidth]{ 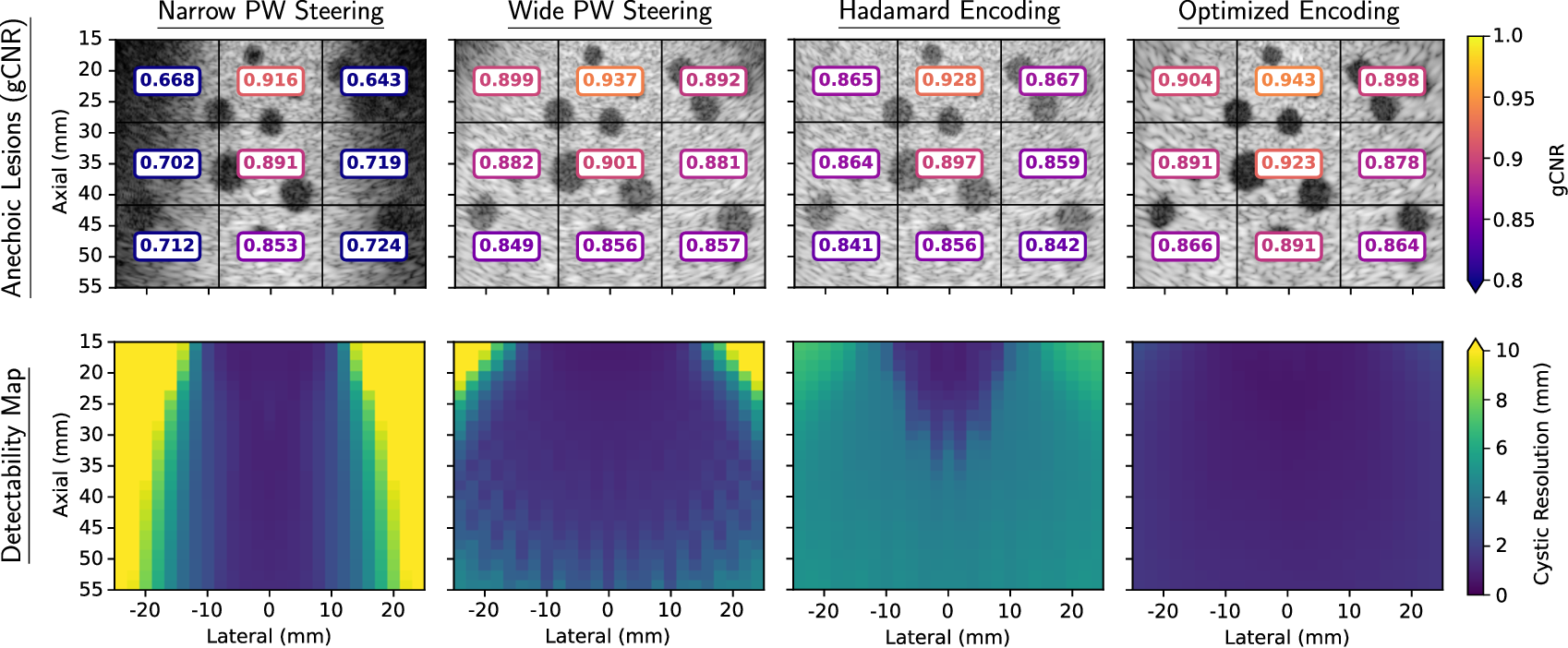}
  \captionsetup{font=small}
  \caption{(top) We plot the average gCNR for anechoic lesions in each region of the viewing range, averaged over randomly located targets over 50 instances of simulated data. Values are displayed for one instance of randomly located targets. (bottom) We plot the cystic resolution throughout the domain. Observe that only our optimized encoding sequence is able to maintain the same quality measured by gCNR and cystic resolution throughout the imaging domain. }
  \label{fig:quantitative_results}
\end{figure}

In contrast to these narrow planewaves, however, our optimal encoding sequence maintains this level of resolution throughout the imaging domain. 
To see this, we use Field~II to simulate the responses from single, isolated scatterers distributed throughout the viewing window, and compute the cystic resolution for each when imaged with the optimized sequence.
Shown in Figure~\ref{fig:quantitative_results}, this procedure generates a map of lesion detectability as a function of target position.
This strongly suggests that the benefits of our optimized encoding sequence stem from more general acoustic principles, rather than overfitting to the training data, as the ML model is never exposed to these isolated point targets during training.

We additionally measure contrast using the gCNR, which also varies according to the spatial position of the lesion target.
To account for this variance, we consider simulated anechoic lesion data for which the position and radius of each lesion is random.
We categorize these lesions based on position, and tabulate the average gCNR for each category in Figure~\ref{fig:quantitative_results}.
To consider more complicated lesion shapes, we also measure the gCNR of the binarized, image-derived data of Figure~\ref{fig:training_data}(b), which we record in Table~\ref{tab:improvement_table_gcnr}.
Altogether, we see that our optimized encoding sequence outperforms conventional planewave and Hadamard encodings in terms of both FOV and contrast.

\begin{table}[htbp]
    \centering
    \begin{tabular}{rcccc}
        \toprule
        \multicolumn{5}{c}{Average gCNR}\\
        \toprule
                                                 & $1^\circ$ Spacing & Full FOV Span & Truncated & Optimized \\
        \multicolumn{1}{c}{\textbf{Target Type}}  & Planewaves        & Planewaves    & Hadamard & Encoding   \\
        \cmidrule(r){1-1} \cmidrule(l){2-5}
         Binarized Image- & \multirow{2}{*}{0.882} & \multirow{2}{*}{0.654} & \multirow{2}{*}{0.888} & \multirow{2}{*}{\textbf{0.891}}  \\
         Derived Contrast &                        &                        &                        &                                  \\\bottomrule
    \end{tabular}
\captionsetup{font=small}
    \caption{We compute the average gCNR over 50 pieces of out-of-sample data whose ground-truth contrast is derived from a grayscale image. Observe that our optimized encoding results in the highest gCNR, emphasized in bold.}
    \label{tab:improvement_table_gcnr}
\end{table}

\begin{figure}
  \centering
  \includegraphics[width=\textwidth]{ 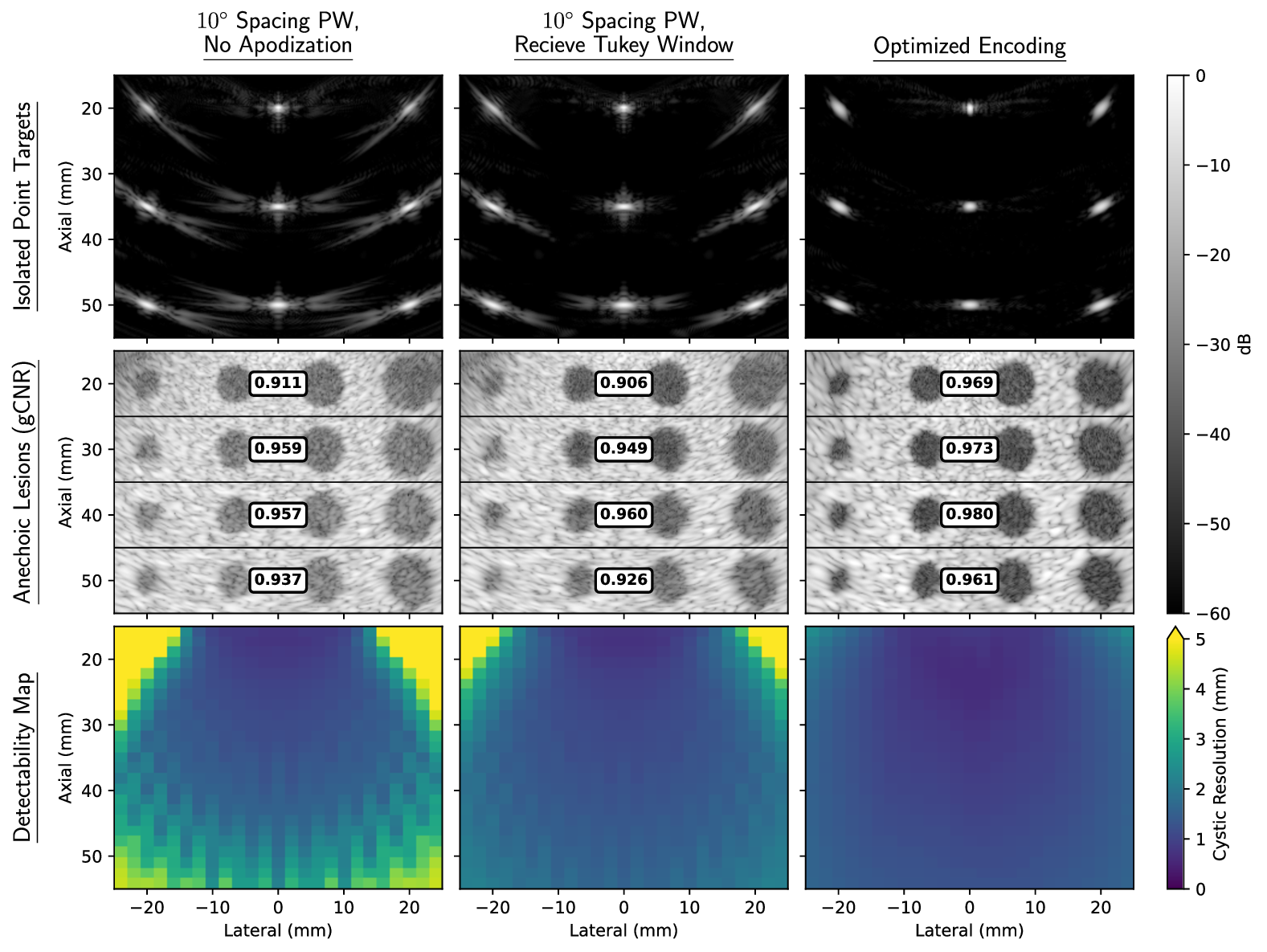}
  \captionsetup{font=small}
  \caption{We compare our technique to a classical application of apodization. We do this visually (top), through the average gCNR across each row of lesions (middle) and through the cystic resolution (bottom). We observe that although the Tukey window does improve the PSF, there is greater improvement from our optimized sequence.}
  \label{fig:full_apodization_example}
\end{figure}

We also compare our optimized sequence to standard techniques in apodization, which similarly seek to improve the PSF by reducing side lobes~\citep{jeong-apodization-2020}.
To do this, we consider planewaves with $10^\circ$ spacing, which offers good resolution at depth while imaging the full FOV, and apply a Tukey window to the receive channel.
As expected, this improves cyst detectability in the image considerably.
However, it does so at the cost of resolution, as shown in Figure~\ref{fig:full_apodization_example}.
Specifically, there are still clear artifacts around point scatterers, and the striation in the cystic resolution map for the unapodized case is still present after apodization.
These effects are not present in our optimized encoding sequence.

\subsection{Restricted Optimization to Either Time Delays or Apodization Weights}\label{sec:only}

We can also apply the proposed ML model and training framework to more restrictive hardware specifications.
In doing so, we demonstrate that there is still considerable improvement present, although such sequences do not perform as well as a fully arbitrary system, as expected.
To this end, we train two separate configurations of the ML model.
In the first, the apodization weights are fixed to some uniform value, and the time delays are the \textit{only} trainable parameters of the ML model.
In the second, only the apodization weights can be changed, and each time delay is fixed to zero in the ML model.
As in Section~\ref{sec:both}, we train the model on anechoic lesions in underdeveloped speckle and test the resulting sequence on data far outside the training set.
We compare the two resulting optimized sequences to the appropriate conventional alternative: wide spanning planewaves for delay-only optimization, and a truncated Hadamard encoding for weight-only optimization.

The results of this comparison are compiled in Table~\ref{tab:improvement_table_loss_type}.
We see that in each case the optimized sequence performs better according to the $\ell^2$ loss metric than its conventional alternative, which we have now shown correlates with high quality as measured by gCNR and cystic resolution.
Furthermore, we observe that while there is still considerable improvement when only delays are optimized through the training of our ML model, the magnitude of improvements when only optimized weights are generated is comparable to that of a fully optimized encoding sequence.
This demonstrates that in this imaging scenario, optimization of the weights has a more significant influence on the quality of the resulting sequence than delay optimization alone.

\begin{table}[htbp]
    \centering
    \begin{tabular}{rccccc}
        \toprule
        \multicolumn{6}{c}{Average $\ell^2$ Loss against Ground-Truth Contrast Map}\\
        \toprule
        & \multicolumn{2}{c}{\textbf{Delay Optimization}} & \multicolumn{2}{c}{\textbf{Weight Optimization}} & \textbf{Both} \\ 
         \cmidrule(lr){2-3} \cmidrule(lr){4-5} \cmidrule(l){6-6}
                                                 & \multirow{2}{*}{Optimized}      & Full FOV               & \multirow{2}{*}{Optimized}      & Truncated                       & \multirow{2}{*}{Optimized} \\
        \multicolumn{1}{c}{\textbf{Target Type}} &                                 & Planewaves             &                                 & Hadamard                        &                            \\ 
        \cmidrule(r){1-1} \cmidrule(lr){2-3} \cmidrule(lr){4-5} \cmidrule(l){6-6}
        Isolated                                 & \multirow{2}{*}{\textbf{16.93}} & \multirow{2}{*}{20.03} & \multirow{2}{*}{\textbf{14.53}} & \multirow{2}{*}{33.87} & \multirow{2}{*}{\textbf{8.129}} \\
        Point Targets                            &                                 &                        &                                 &                        &                                 \\
        \cmidrule(r){1-1}
        Underdeveloped                           & \multirow{2}{*}{\textbf{310.2}} & \multirow{2}{*}{331.6} & \multirow{2}{*}{\textbf{303.3}} & \multirow{2}{*}{366.3} & \multirow{2}{*}{\textbf{252.9}} \\
        Speckle                                  &                                 &                        &                                 &                        &                                 \\
        \cmidrule(r){1-1}
        \multirow{2}{*}{Anechoic Lesions}        & \multirow{2}{*}{\textbf{231.2}} & \multirow{2}{*}{250.5} & \multirow{2}{*}{\textbf{217.1}} & \multirow{2}{*}{257.3} & \multirow{2}{*}{\textbf{205.2}} \\
                                                 &                                 &                        &                                 &                         &                                 \\
        \cmidrule(r){1-1}
        Image-Derived                            & \multirow{2}{*}{\textbf{165.7}} & \multirow{2}{*}{166.8} & \multirow{2}{*}{\textbf{164.2}} & \multirow{2}{*}{174.4} & \multirow{2}{*}{\textbf{160.8}} \\
        Contrast                                 &                                 &                        &                                 &                        &                                 \\
        \cmidrule(r){1-1}
        Binarized Image-                         & \multirow{2}{*}{\textbf{243.3}} & \multirow{2}{*}{258.8} & \multirow{2}{*}{\textbf{241.2}} & \multirow{2}{*}{274.7} & \multirow{2}{*}{\textbf{230.1}} \\
        Derived Contrast                         &                                 &                        &                                 &                        &                                 \\\bottomrule
    \end{tabular}
  \captionsetup{font=small}
    \caption{We consider training ML models whose trainable parameters are restricted to only a single component of the encoding sequence. In both cases, we see that the per-component optimized encoding sequence outperforms the conventional alternative on each type of imaging target.}
    \label{tab:improvement_table_loss_type}
\end{table}

%% file: experimental_results.tex
\subsection{Optimization in the Presence of Noise}

These simulated results demonstrate the theoretical utility of applying an ML optimization strategy for data acquisition and imaging within the REFoCUS framework.
However, we can also verify that our optimized sequences produce meaningful improvements in several practical scenarios.
In the first, we consider a separate configuration of the ML model that is trained to emulate an imaging system under the influence of electronic noise.
We recreate this during training by the addition of random noise to the encoded channel data prior to imaging.
In the current example, this additive noise is sampled from a random normal distribution, filtered to have a bandwidth equal to that of the transducer, and then scaled to produce an average of 20 dB channel signal-to-noise ratio. 
It is in this context that restricting the set of apodization weights to an $\ell^\infty$ ball is most critical, as otherwise all noise would be suppressed simply by arbitrarily increasing the apodization weights.

\begin{figure}[htbp]
  \centering
  \includegraphics[width=\textwidth]{ 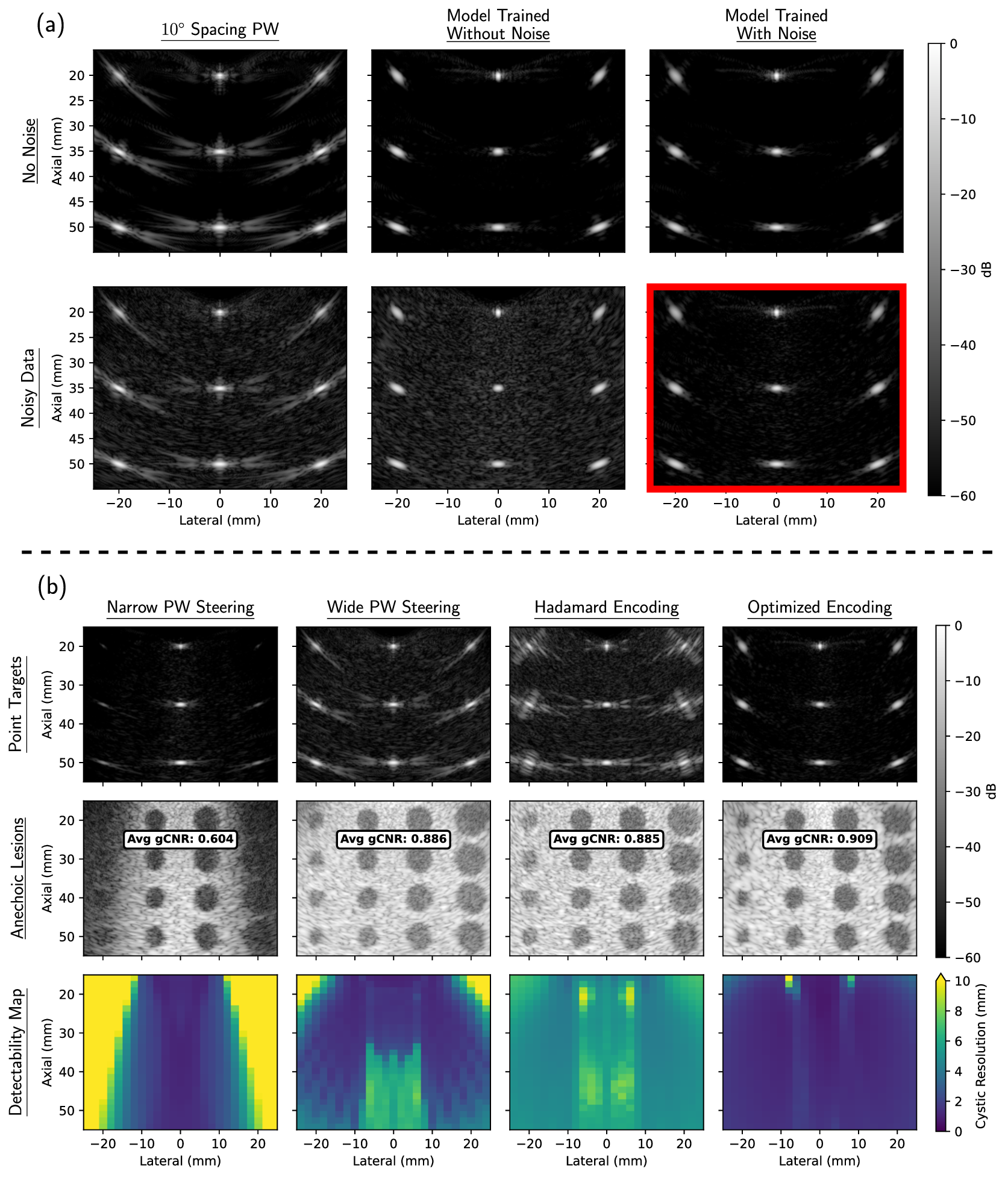}
  \captionsetup{font=small}
  \caption{Effects of noise on training/evaluation of an ML model. (a) We compare encoding sequences generated by different ML models trained with and without the presence of electronic noise. (b) We make additional comparisons to the case emphasized in red in (a), comparing conventional transmit sequences on noisy RF data with a sequence optimized in the presence of additive noise. We see a greater qualitative suppression of noise (top), as well as improvements in gCNR (middle) and the point spread function as measured by the cystic resolution (bottom). }
  \label{fig:total_noise}
\end{figure}

By accounting for this noise during training, the ML model generates a new encoding sequence that specifically suppresses this noise.
We see this in comparison to the optimized encoding sequence evaluated throughout Section~\ref{sec:both}.
In Figure~\ref{fig:total_noise}(a), we consider these two sequences, trained with and without the presence of noise, along with a conventional planewave configuration for comparison.
We then apply each sequence to data to which noise is, or is not added.
In each of the four cases, we see considerable improvement over the conventional planewave approach.
However, we can also see that these improvements are most prominent when the ML model is trained with the same type of noise present during evaluation, the case emphasized in Figure~\ref{fig:total_noise}(a) in red.
In Figure~\ref{fig:total_noise}(b), we consider a more complete evaluation of the case where noisy data is imaged using a sequence trained with the same noise model.
There is the same degree of improvement as the noiseless case of Figure~\ref{fig:total_improvements}, featuring dramatic improvements in the gCNR and cystic resolution.

\subsection{Experimental Verification}\label{sec:experimental}

Finally, we verify these results by directly implementing the optimized encoding sequence discussed in Section~\ref{sec:both} in physical hardware.
In the following experiments, we use the Verasonics Vantage 256 research scanner (Verasonics, Inc., Kirkland, WA) with the P4-2v phased array transducer (3 MHz transmit frequency, 64 elements, 0.3 mm pitch, 10V transmit voltage).
Received channel data were stored for processing offline, where they were decoded and used to produce an image in the same manner as in simulation.
For comparison, we use sequences of 15 transmits that generate planewaves of varying span (15, 60, and 150 degrees). 
To reproduce the effects of arbitrarily scaled weights (rather than restricting to $\pm1$ through pulse inversion) we use the Verasonics apodization function to scale the duty cycle of the transmit excitation so that it matches the output amplitude.

\begin{figure}[htbp]
  \centering
  \includegraphics[width=\textwidth]{ 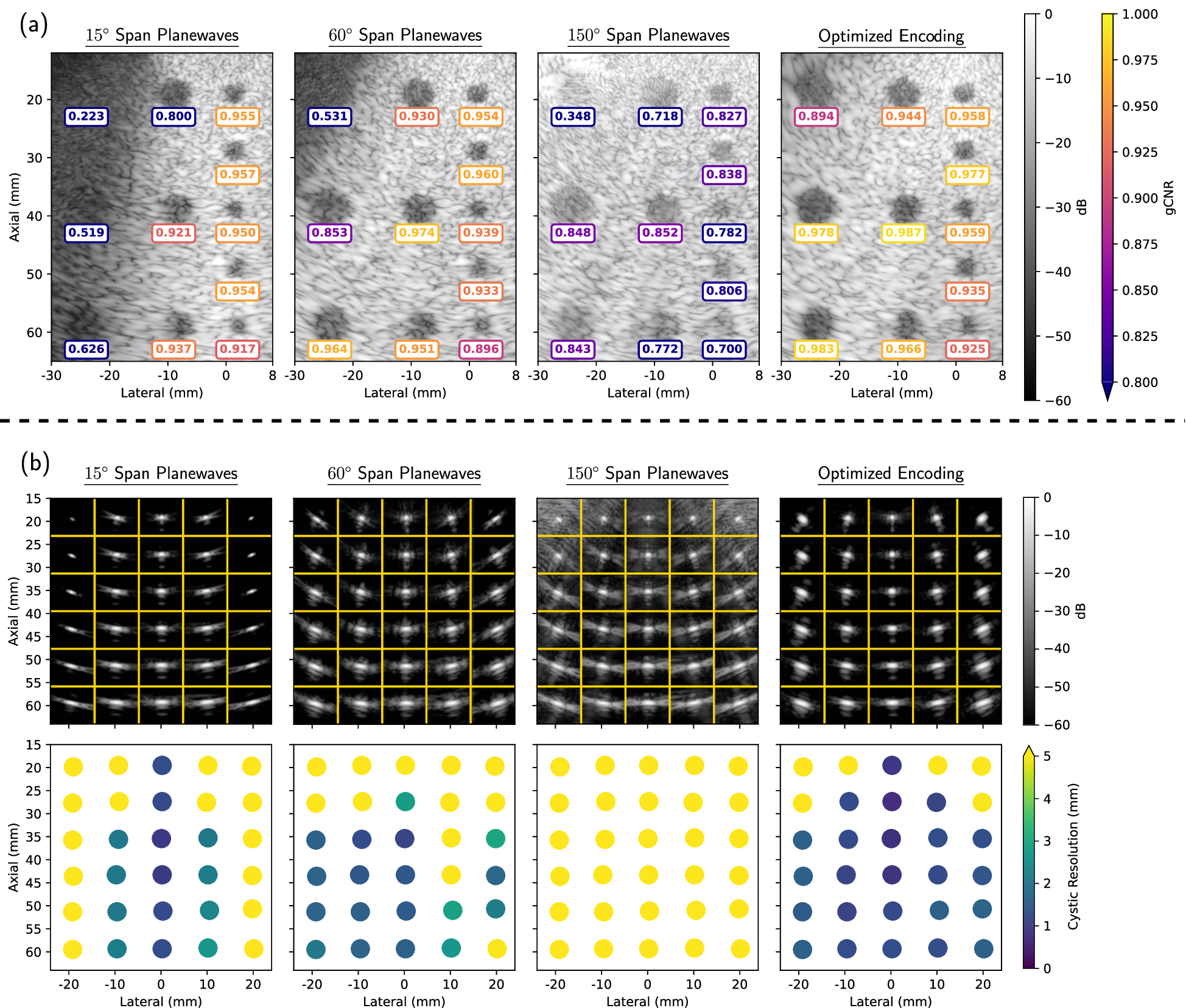}
  \captionsetup{font=small}
  \caption{Experimental validation of simulated results. (a) We image the ATS 539 multi-purpose phantom using different encoding sequences. Although the ML model is optimized for a symmetric FOV, the fixed lateral position of the transducer was intentionally placed off-center to demonstrate improved FOV, while fully capturing each target in the asymmetric phantom. The average gCNR over seven realizations of the speckle pattern is shown beneath each anechoic lesion. (b) We create a montage of 30 individual point target images. Each is obtained by translating the P4-2v transducer around a fixed wire target in a water tank. To provide an appropriate comparison between different targets within the montage, each image is displayed with the same amount of gain (top). We use these images to compute the cystic resolution throughout the domain (bottom).}
  \label{fig:total_experiments}
\end{figure}

The first experiment utilizes an ATS 539 multi-purpose phantom (CIRS, Inc., Norfolk, VA) to confirm the ability of optimized encoding sequences to improve imaging on tissue-like material.
Within the phantom we image anechoic lesions of varying position and radius in background speckle, and compute the gCNR of each with respect to surrounding background speckle.
We show these gCNR values below each lesion in Figure~\ref{fig:total_experiments}(a).
For a fixed lateral position, we acquire data using each encoding sequence at each of seven elevation positions of the phantom.
In this way, we image anechoic lesions in the same positions with different realizations of background speckle, and compute the average gCNR across all realizations.

We can see that these results match closely with our numerical simulations, in that our optimized sequence is able to produce the depth of focus and resolution of a very narrow span of planewaves, while still maintaining a full FOV throughout the range.
Importantly, these improvements persist beyond the fixed viewing window of the training data, which extends only to a depth of 55 mm.
This further emphasizes that our encoding sequence has not simply specialized to the specific characteristics of the training data.

We can also recreate our simulated point target images, which we can in turn use to measure the cystic resolution throughout the imaging domain (albeit on a much coarser grid).
To accomplish this, we image a custom, single target wire phantom (0.03 mm tungsten wire) in a water tank.
Using the AIMS III Hydrophone Scanning System~(Onda Corporation, Sunnyvale, CA), we mechanically translate the suspended P4-2v transducer around the fixed wire so that the target can be imaged from a 6$\times$5 grid of target positions ($\sim$10 mm lateral spacing, $\sim$8 mm axial spacing).

Although we only ever image one point target at a time with this experimental setup, in Figure~\ref{fig:total_experiments}(b), we present a \textit{montage} of these images distributed appropriately in space. 
To ensure a fair visual comparison of each point target, we apply a \textit{uniform} level of gain to each image in the domain.
This allows us to very clearly see the effect of our optimized sequence on the FOV of the imaging system, where points in the upper corners of the viewing window are either undetectable or suffer from very poor resolution.
In contrast, the optimized encoding sequence provides a uniform degree of resolution at all points in the domain. 

Furthermore, we can see that the off-axis scattering artifacts present are greatly reduced when imaged with the optimized sequence. 
As before, we quantify this effect using the cystic resolution, which we can compute for each point target in Figure~\ref{fig:total_experiments}(b).
Although our ability to perfectly capture this metric is slightly hampered by irregularities in the experimental setup, we still observe near uniform improvement over each conventional sequence, analogous to results observed in simulation.

%% file: discussion.tex
\section{Discussion}
Our machine learning framework represents a novel method of generating transmit sequences that ultimately result in higher quality images than current standards.
However, as with most machine learning applications, maximizing the efficacy of the optimized sequence requires careful consideration of the ML model configuration.
For example, one must select a number of standard ML hyperparameters, such as descent algorithm, batch size, learning rate, etc.
Furthermore, this particular usage of ML in ultrasound imaging brings about additional, unique considerations.
In this section, we discuss our own exploration of these considerations, and make observations that we believe will be relevant in any future ultrasound ML application that directly incorporates a beamformer into the architecture.

\subsection{Significance of the Initial Condition}
Because the minimization problem presented in Equation~\ref{eqn:optimized_sequence} is not convex, there is no expectation that any optimization procedure can find the global minimum, should one even exist.
Instead, the particular local minimum approached by our optimization procedure is highly dependent on the initial condition.
Because locally optimal encoding sequences vary in quality, it is important to select an initial condition that is at or near a good local minimum.
The interpretability of the parameters of our ML model offers a distinct advantage over conventional deep learning applications in accomplishing this, as we can intelligently explore the effects of different initial conditions.

\begin{figure}[htbp]
  \centering
  \includegraphics[width=\textwidth]{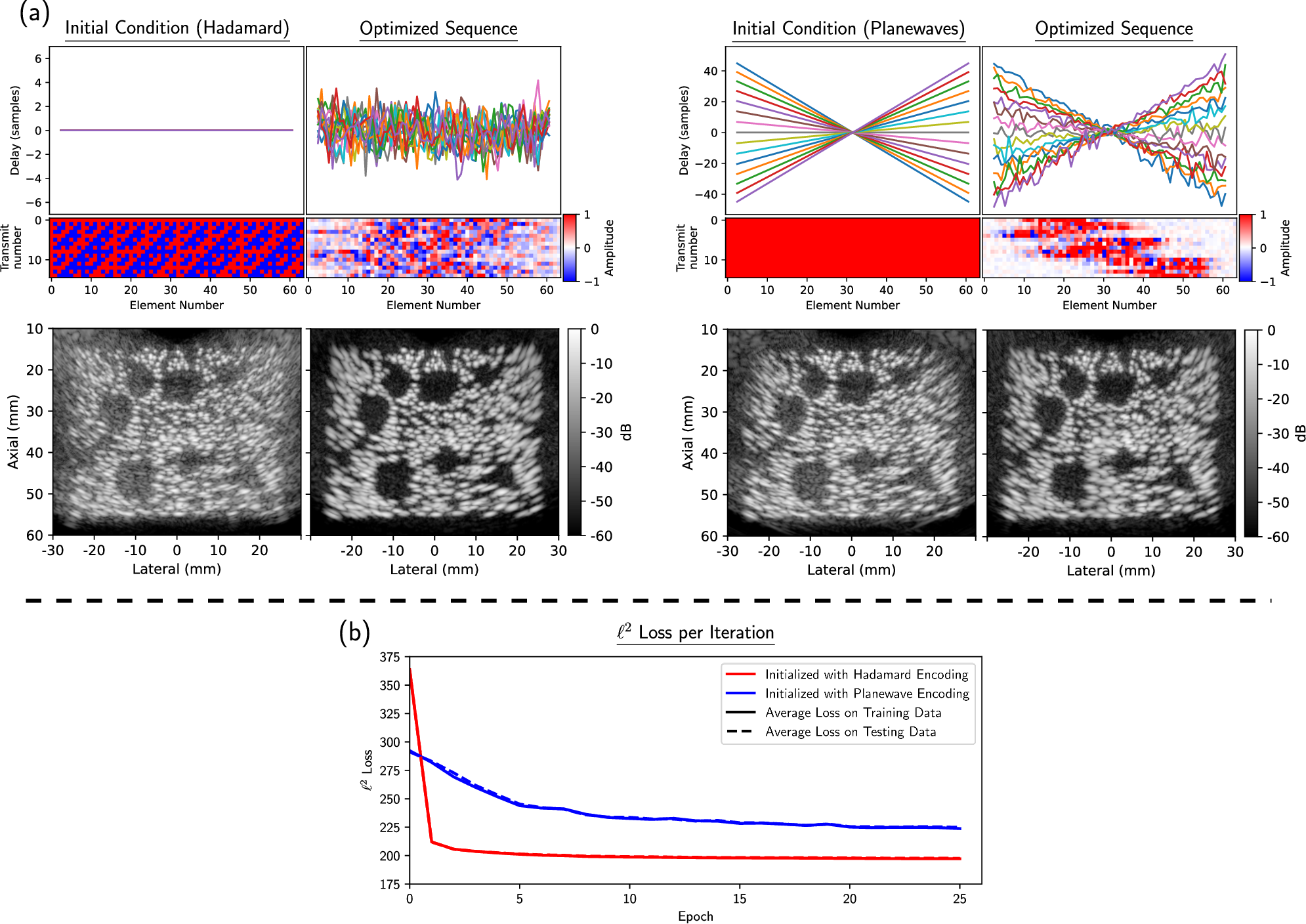}
  \captionsetup{font=small}
  \caption{A comparison between two different encoding sequences used as initial conditions. (a) For each, we plot the transmit sequence before and after optimization, along with one piece of testing data for each. In the delay plots, each transmission is represented by a single line. (b) For each initial condition, we plot the per-epoch $\ell^2$ loss function averaged over the training and testing data sets.}
  \label{fig:total_model_improvements}
\end{figure}

For example, an instance of our ML model initialized with a planewave encoding will converge to transmissions that are steered in the same way, albeit with less well-defined wavefronts.
On the other hand, initializing with a truncated Hadamard code causes the parameters to converge to a sequence with no visible steering.
We can see these qualitative behaviors in Figure~\ref{fig:total_model_improvements}(a).
Despite this, we have found experimentally that a truncated Hadamard sequence is the superior initial condition, despite being a generally lower quality transmission sequence.
This can be seen through the $\ell^2$ loss plotted over the course of training epochs for each initial condition in Figure~\ref{fig:total_model_improvements}(b).
This can be explained by the conventional Hadamard sequence allowing the ML model to begin with near-optimal FOV, as well as both positive and negative apodization weights.
Except for cases in which delays or weights must be fixed as in Section~\ref{sec:only}, each optimized encoding sequence used in this paper was generated by an ML model with this configuration as the initial condition.

In our exploration of this topic, we have found that recovering a known, conventional sequence is exceptionally difficult except in the simplest toy problems.
For example, although planewave transmissions are generally performant, the ML model cannot identifiably recover the same geometric properties that make them desirable.
On the other hand, the optimal characteristics of the resulting sequence are quite opaque: while optimized sequences appear to be a simple perturbation of the initial condition, performing this perturbation randomly fails to achieve the same improvements. 
This suggests that it would not be possible to discover these sequences \textit{without} the use of the proposed ML model.

\subsection{Significance of the Training Data}\label{sec:significance_training_data}
In Figure~\ref{fig:total_model_improvements}(b), we also observe that for either initialization, our ML model performs nearly identically between our training and testing data.
This is a highly desired property in this context, as ultrasound imaging can be theoretically formulated as a convolution of a spatially varying point spread function across each scatterer in the domain, the results of which are then summed to form an image.
Therefore, an effective manner of optimizing image quality for arbitrary data would be to optimize this PSF itself, concentrating it at zero offset~\citep{goudarzi-deep-reconstruction-2022}.

Although our loss function does not reference features of the PSF directly, our choice to use underdeveloped speckle as training data implicitly encourages the same type of improvement at a lesser computational burden.
This is in part because the scatterers within each sample are spread out enough that their off-axis scattering artifacts do not necessarily overlap one another, as would be the case with standard background speckle.
This means the $\ell^2$ loss function can only be meaningfully decreased by acoustically concentrating the energy of each scatterer inwards, thereby improving the PSF.
At the same time, scatterers across \textit{all} samples in the training data are densely distributed throughout the imaging domain.
This means that learning improvements that span the entire FOV can be obtained with relatively few samples of type of training data.

\subsection{Significance of the Imaging Target}\label{sec:imaging_target}
\begin{figure}[htbp]
  \centering
  \includegraphics[width=\textwidth]{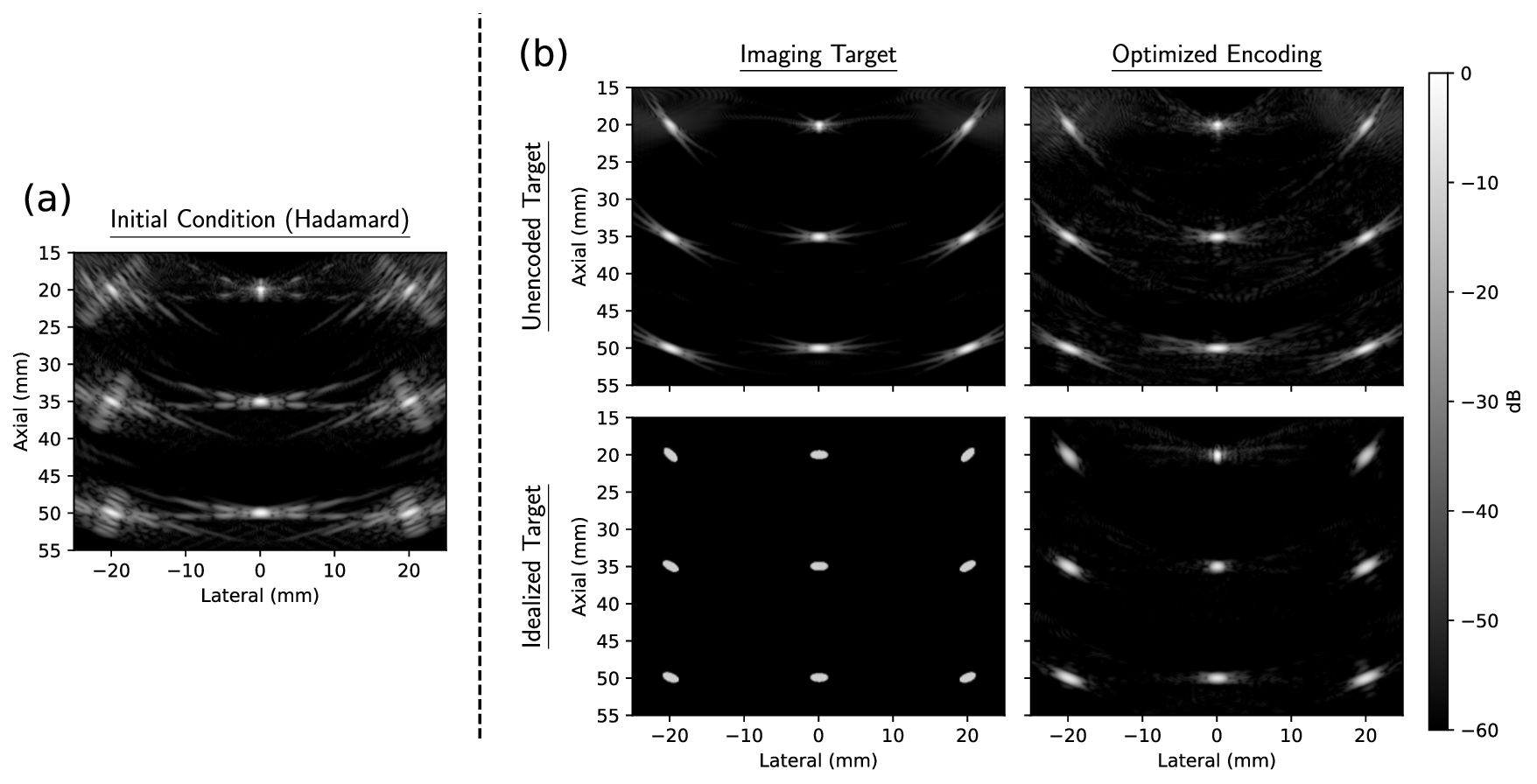}
  \captionsetup{font=small}
  \caption{The quality of the encoding sequence depends heavily on the imaging target used during training. (a) We initialize two ML models identically. (b) When the imaging target of the $\ell^2$ loss is unencoded data (top), the resulting sequence has the same artifacts present in the beamformed image. By using an idealized imaging target, i.e., one that represents the ground-truth contrast, the resulting sequence can suppress these artifacts (bottom).}
  \label{fig:training_target_results}
\end{figure}

Although we exclusively use a loss function that compares produced B-mode images with some ground-truth imaging target, there is still much flexibility within that choice.
As stated in Section~\ref{sec:acquisition_training_data}, a natural choice for the imaging target is an image derived from ground-truth multistatic data.
While this produces target images without any encoding artifacts, we have found that an ML model trained with these targets produces poor image quality relative to alternatives.
This is because DAS beamforming with even unencoded multistatic data results in a PSF with undesirable features such as the visible side lobe artifacts present throughout the domain.
When the ML model is trained on such target images, the transmit sequence is implicitly encouraged to accurately reconstruct these artifacts.
Instead, we see that training against the exact ground-truth contrast produces images with a higher quality PSF, as seen in Figure~\ref{fig:training_target_results}.

\begin{figure}[htbp]
  \centering
  \includegraphics[width=0.8\textwidth]{ 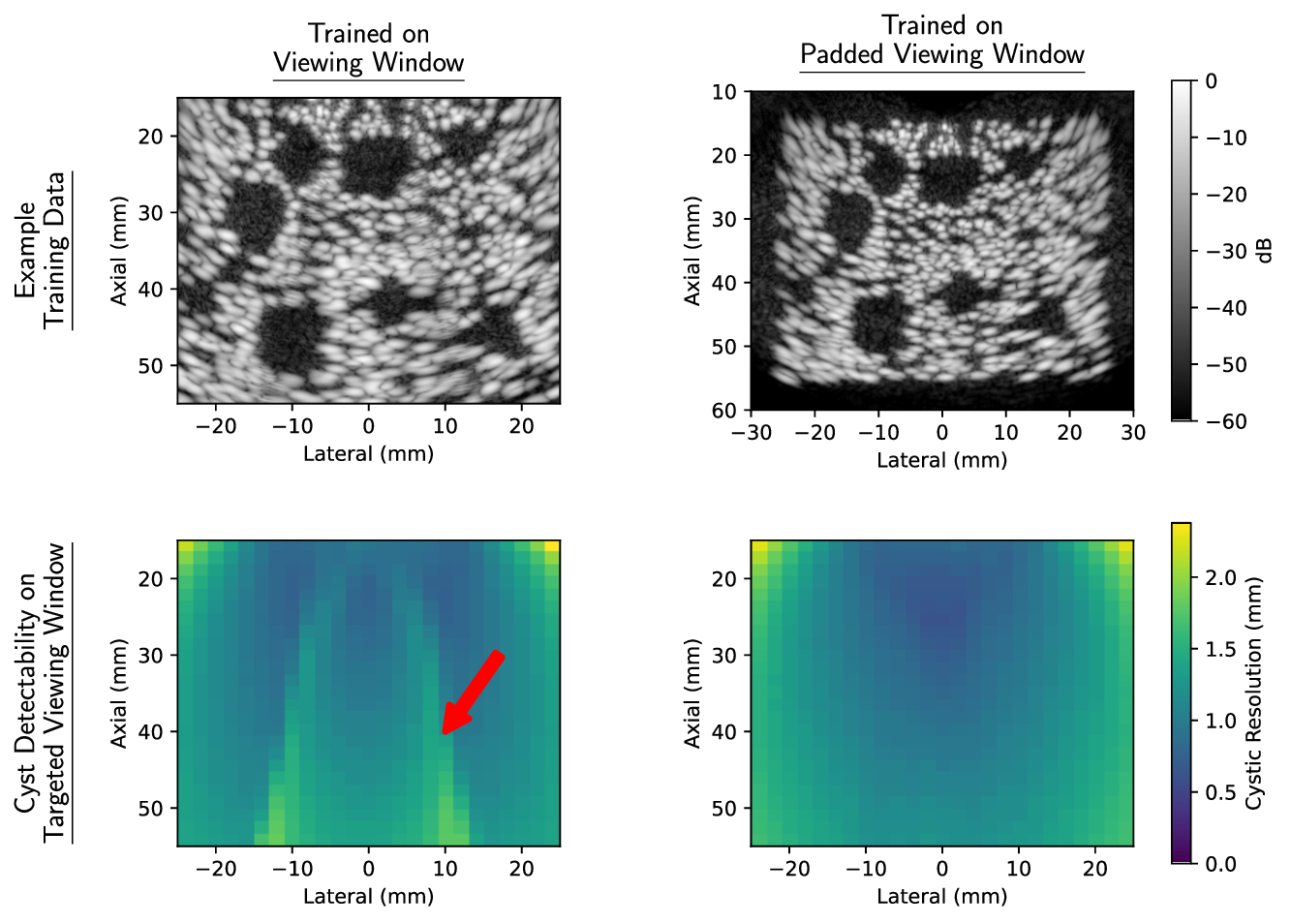}
  \captionsetup{font=small}
  \caption{The imaging target must be carefully selected to ensure there is not an unnecessary dependence on the specifics of the imaging domain. For example, by expanding the window during training (right), we are able to deal with streaking artifacts (indicated with an arrow) that arise from unsuppressed scattering just beyond the imaging domain.}
  \label{fig:cystic_resolution_grid_padding}
\end{figure}

Additionally, we see the most uniform improvement in image quality when we ``pad'' images of underdeveloped speckle so that the final viewing window is itself located in an anechoic field.
This ensures that the encoding sequence is not unnecessarily dependent on the specific domain geometry. 
The consequences of this are best seen through the plots of cystic resolution in Figure~\ref{fig:cystic_resolution_grid_padding}, where the first encoding sequence is trained on images whose extent is exactly that of the final viewing window. 
In contrast, the second encoding sequence is trained on images located in a slightly wider anechoic void. 
As we can see, adding this empty space causes the PSF to more smoothly vary throughout the domain, and failing to do so results in additional artifacts for point targets near the center of the region.
In effect, these streaks depict positions for which scattering artifacts extend beyond the imaging domain, and the introduced error is not captured by the loss function. 

\subsection{Significance of the Loss Function}\label{sec:loss_function}

By using our custom implementation of a differentiable beamformer, we are able to train our ML model according to improvements in an \textit{imaging} metric, which is a novel departure from existing work that only considers the recovery of the multistatic data set.
Importantly, we can demonstrate the existence of transmit sequences that produce images with higher resolution and contrast, even though the encoding itself produces a \textit{worse} reconstruction of $\CU$ from $\CS$.

This is primarily because a perfect reconstruction of $\CU$ will only produce the original unencoded image, and this type of image lacks many desirable properties, as discussed in Section~\ref{sec:imaging_target}.
To create a concrete example of this phenomenon, we compare to an ML model within our framework that is instead configured to directly optimize the recovery of the multistatic data set by minimizing

\begin{align}\label{eqn:STAloss}
    \mathcal{L}_{\text{STA}}(\widehat{\CU}; \CU) := \frac{1}{N_TN_RN_\omega}
    \big\|
    \CU - \widehat{\CU}
    \big\|^2_2\,.
\end{align}

\begin{figure}[htbp]
  \centering
  \includegraphics[width=0.8\textwidth]{ 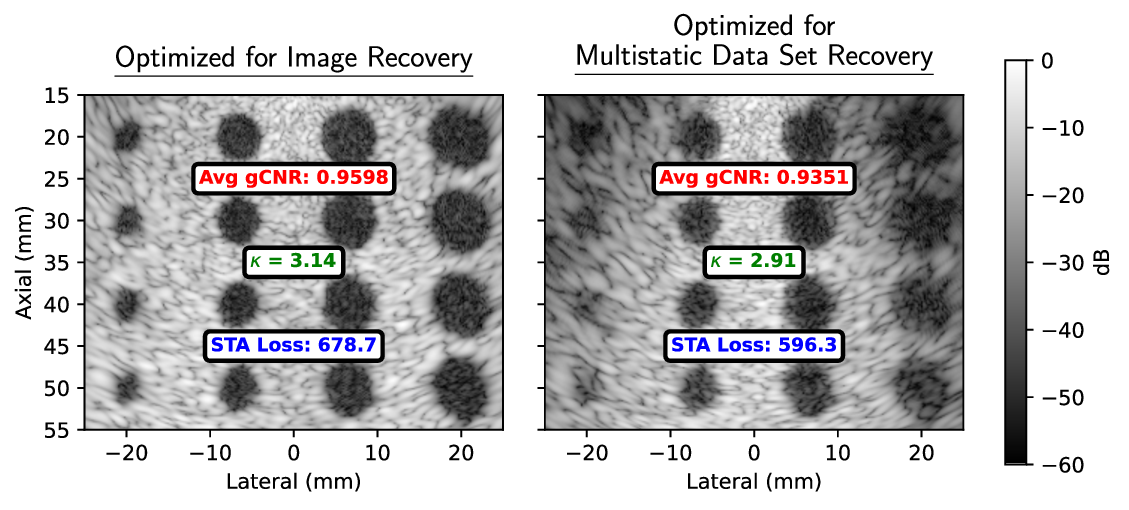}
  \captionsetup{font=small}
  \caption{Comparison of our optimized sequence (left) to one that is trained to optimize reconstruction of the multistatic data set (right). In spite of a worse multistatic reconstruction measured by the STA loss and comparable condition number $\kappa$ for the encoding matrices, our strategy for optimization produces visibly improved contrast and resolution throughout the image.}
  \label{fig:sta_comparison}
\end{figure}

Consider Figure~\ref{fig:sta_comparison}.
We see that this new configuration successfully generates an encoding sequence that produces a lower STA loss than the previously considered encoding sequence optimized for image recovery.
Yet despite this numerical improvement, there is a clear superiority in image quality when the optimization process considers image formation, leading to improvements both visually and in terms of the gCNR averaged over each lesion.

We take this as a counterexample to the common convention that a better conditioned encoding necessarily corresponds to higher quality images.
To account for the effects of Tikhonov regularization, we measure the condition number of the collection of encoding matrices $\CH$ as $\kappa = \| \CH_0 \| \cdot \| \CH^\dagger_0 \|$, doing so because the encoding matrix for the lowest frequency $\CH_0$ has the poorest conditioning among the collection~$\CH$.
We see that both encoding sequences bring about reasonably well-conditioned matrices following Tikhonov regularization, but there are still clear qualitative differences between the two images.

%% file: conclusions.tex
\section{Conclusions}
In summary, we have shown that the principles of machine learning can be used to generate encoding sequences that improve the quality of ultrasound imaging within the context of the REFoCUS model.
An important theoretical consequence of our ML procedure is its ability to discover currently unknown transmit sequences with desirable properties. 
For example, the various ML models trained for demonstration throughout this paper have each found transmit sequences that are ``high-quality'' in a sense that is acoustically meaningful, yet is not directly encouraged by the training procedure.
That is to say, although the sequence is selected so that it minimizes one particular $\ell^2$ imaging metric, it ultimately improves quality along a number of other metrics across disparate classes of data.
At the same time, the exact acoustic properties that these sequences share remains obscure.
This means that our ML model has stumbled upon a currently unexplored regime of transmit sequences, and there remain unanswered questions as to the unifying features of such sequences.

By using an ML model whose parameters are exactly those of an encoding sequence, this approach offers a high degree of flexibility to different imaging scenarios, making it an important foundation for future endeavors.
For example, the high degree of generalizability shown in Figure~\ref{fig:total_improvements} indicates that, in sharp contrast to other deep-learning image formation and analysis tasks, it may be possible to train with a limited amount of \textit{in vivo} data without a severe degradation in image quality.
The theoretical flexibility of the REFoCUS framework is a key component of this type of investigation, as it allows for uniform treatment of highly variable kinds of transmit sequences.
Beyond the proposed framework, we wish to further explore the capabilities of machine learning within REFoCUS, motivated by our ability to incorporate beamforming as a layer of an ML architecture, potentially with its own set of trainable parameters.
As has been the case in this work, such technology will ultimately allow us to continue developing techniques that directly improve image quality for applications of interest.

\section{Data Availability}
All data that support the findings of this study are included within the article (and any supplementary information files).